\documentclass[letterpaper,twocolumn,10pt]{article}
\usepackage{usenix}

\usepackage{00_preamble}
\usepackage{00_macros}

\begin{document}

\title{\LARGE \bf Security-Aware Sensor Fusion with MATE: the Multi-Agent Trust~Estimator}

\newif\ifAnonymize

\Anonymizefalse

\ifAnonymize

\else

    \author{R. Spencer Hallyburton and Miroslav Pajic%
    \thanks{This work is sponsored in part by the ONR under agreement N00014-23-1-2206, AFOSR under award number FA9550-19-1-0169, and by the NSF under CNS-1652544 award and the National AI Institute for Edge~Computing Leveraging Next Generation Wireless Networks, Grant CNS-2112562.}
    \thanks{R. S. Hallyburton and M. Pajic are with Department of Electrical and Computer Engineering, Duke University, Durham, NC 27708, USA;
    {\tt\small \{spencer.hallyburton,~miroslav.pajic\}@duke.edu}.}%
    }
\fi

\maketitle

\begin{abstract}   
    Lacking security awareness, sensor fusion in systems with multi-agent networks such as \emph{smart cities} is vulnerable to attacks. To guard against recent threats, we design \emph{security-aware sensor} fusion that is based on the estimates of distributions over \emph{trust}. Trust estimation can be cast as a hidden Markov model, and we solve it by mapping sensor data to trust \emph{pseudomeasurements} (PSMs) that recursively update trust posteriors in a Bayesian context. Trust then feeds sensor fusion to facilitate trust-weighted updates to situational awareness. Essential to security-awareness are a novel field of view estimator, logic to map sensor data into PSMs, and the derivation of efficient Bayesian updates. We evaluate security-aware fusion under attacks on agents using case studies and Monte Carlo simulation in the physics-based Unreal Engine simulator, CARLA. A mix of novel and classical security-relevant metrics show that our security-aware fusion enables building trustworthy situational awareness even in hostile conditions.
\end{abstract}
\section{Introduction}

Securing cyber-physical systems (CPS) against attacks is a perennial problem. The proliferation of low-cost, networked agents has upended traditional monolithic platforms and elucidated security challenges. Of particular concern is \emph{collaborative sensor fusion}. While multi-agent fusion is effective at maintaining accurate situational awareness (SA), it is vulnerable to adversaries as it assumes agents behave according to benign models of sensing. Even small numbers of compromised nodes can take down collaborative data fusion in large-scale heterogeneous networks~\cite{lo2007illusion,hallyburton2024bayesian}. 

In this work, we consider an urban environment composed of networked mobile and static agents (``\emph{smart city}'') that wish to maintain accurate SA by estimating both the number of objects that exist and the states of those objects (e.g.,~position, velocity) with \emph{multiple target tracking} (MTT). Unfortunately, agents are limited by inter-object occlusions and may be adversarially compromised. To overcome such limitations, agents are connected to the smart city network and perform \emph{collaborative sensor fusion}. Agents send local information to the smart-city \emph{data aggregator} that builds unified SA on the urban environment with MTT. Collaboration is useful for distributing both global SA and specific warnings including intersection collision warnings (ICW), mobile accessible pedestrian signaling (MAPS), and blind merge warnings (BMW) during safety-critical interactions~\cite{ansari2021v2x}.

CPS security necessitates measures both \emph{proactive} at the network-level (e.g.,~asymmetric cryptography) and \emph{reactive} at the algorithm-level (e.g.,~misbehavior detectors, MBDs)~\cite{kwon2014proactive,yue2007intrusion}. We consider an \emph{insider} attacker possessing valid cryptographic keys. In each frame, the adversary may perturb sensing/detection data of the compromised agent by injecting false positives (FPs), creating false negatives (FNs), or translating existing objects. In the worst case, manipulations will propagate over time in attacker-defined trajectories stealthily consistent with plausible dynamics to engender safety-critical outcomes such as an impending collision.

Since the set of compromised agents is unknown, security awareness at data fusion is important. Unfortunately, many secure computing algorithms were designed with Byzantine fault models for distributed ad-hoc networks, e.g.,~\cite{theodorakopoulos2004trust,wang2006autonomous}, are MBDs that only detect malicious data and cannot recover trusted SA~\cite{wang2006autonomous,golle2004detecting}, or require that the ego agent is trusted~\cite{wang2006autonomous,theodorakopoulos2004trust}. These have limited applicability for autonomous agents with \emph{error-prone} perception data derived from real sensors in applications such as smart cities or remote reconnaissance.

Consequently, in this work, we design a two-part \emph{security-aware sensor fusion} algorithm consisting of (1) estimating probability density functions (PDFs) over the \emph{trust} of agents and tracked objects, and (2) trust-informed collaborative data fusion. The trust of agents and their tracked objects are estimated in a Bayesian context allowing incorporating prior information and measurement uncertainty. Trust-informed data fusion then influences the multi-platform data aggregator. The security-aware sensor fusion both detects misbehaving~agents \emph{and} recovers accurate SA under adversarial manipulation.

Trust estimation is a two-step hidden Markov model (HMM). The first step is to propagate the estimate forward in time. The second step is to update the estimate with measurements. Since there is no sensor providing direct measurements of trust (unlike e.g.,~GPS providing position), we design a novel method of mapping real perception-oriented sensor data to trust \emph{pseudomeasurements} (PSMs) composed of a value and confidence on domain $[0,1]$. We decompose the trust update between agents/objects using conditional probability inspired by Gibbs sampling and employ conjugate prior-likelihood pairs to obtain feasible and efficient updates to trust PDFs. 

PSMs are made comparing an agent's locally-estimated SA with a prediction of what it \emph{should} have observed. The prediction leverages dynamic estimates of the agent's field of view (FOV) at each timestep. Prior works fixed constant FOV shapes (e.g.,~circles~\cite{golle2004detecting,soleymani2017secure,tsukada2022misbehavior,allig2019trustworthiness}). This restriction fails in realistic cases because occlusions by objects/infrastructure constantly alter each sensor's FOV. Instead, to build robust FOV models that are dynamic on a frame-to-frame basis and capture occlusions, we implement a ray tracing algorithm on each agent's LiDAR point cloud to estimate FOVs online. 

Without security-informed fusion, any compromised agent can corrupt collaborative SA, even in large networks~\cite{lo2007illusion,hallyburton2024bayesian}. Instead, we propose trust-informed multi-agent fusion. Agent trust PDFs are used as weights so distrusted agents have less influence on collaborative SA. Object trust PDFs inform track management, as distrusted tracks are flagged for investigation or removed. Employing trust-informed data fusion enables mitigation and recovery from adversarial influence on SA.

Evaluations of attacks and defenses are performed on longitudinal multi-agent scenes constructed from the pipelined smart-city dataset generator~\cite{hallyburton2023datasets} built on the Unreal Engine simulator, CARLA~\cite{2017carla}. We design baseline (unattacked) scenes and derive parameterized adversarial versions by applying threat models from state of the art attacks on sensing (e.g.,~\cite{2019cao-spoofing, 2022hally-frustum, hallyburton2023partial}). We Monte Carlo (MC) randomize the adversary parameters to generate a large volume of attacks for analysis. We convey model performance with detailed case studies highlighting the efficacy of security-aware fusion.

Two traditional classes of metrics capture performance of the object existence (number of objects) and state estimation capability of unsecured and trust-informed fusion: (1) precision, recall, and F1-score, and (2) optimal subpattern assignment (OSPA)~\cite{schuhmacher2008ospa}. To evaluate MATE, we derive novel trust-oriented metrics that compare trust PDFs to agent and track states. Metrics illustrate that security-aware fusion detects and identifies distrusted agents/tracks and recovers accurate SA despite adversarial data manipulation. Compared to fusion lacking security awareness, our approach leads to $94\%$ reduction in adversary-driven OSPA error versus the unattacked baseline. Trust-based metrics report MATE adeptly detects compromised agents with near $90\%$ accuracy. Security-aware fusion maintains high levels of mission performance in both benign and adversarial scenarios and is sufficiently robust to be deployed in dynamic, unknown environments.

\noindent\textbf{Contributions.} \ In summary, our contributions are:
\vspace{-4pt}
\begin{itemize}[leftmargin=12pt]
    \setlength\itemsep{-4pt}
    \item First full-stack framework and implementation for security-aware sensor fusion. Multi-agent trust estimation (MATE) and trust-informed data fusion significantly mitigates adversaries in multi-agent networks.
    \item Real-time LiDAR-based FOV estimation to predict the visible area and observable objects free from occlusions.
    \item Novel trust pseudomeasurement algorithm to map real sensing data to the trust domain.
    \item Efficient updates to agent/track trust PDFs using Gibbs decomposition and weighted conjugate priors.
    \item Case studies and MC trials of benign/adversarial scenes with state of the art threat models. Systematic parameter tuning and evaluation with classical and novel metrics.
\end{itemize}
\section{Background}

Perception and sensor fusion have been the subject of much security research. 
\cite{2014cyberattacks} illustrated physical attacks on sensing spawning derivative works in LiDAR~\cite{2019cao-spoofing, 2022hally-frustum} and camera-based~\cite{thys2019fooling, eykholt2018robust} security for AVs.~\cite{hallyburton2023partial} discussed partial-information attacks compromising only a subset of sensors.

\emph{Proactive} security (e.g., public key infrastructure, PKI) reduces access to malicious outsiders while \emph{reactive} security handles insiders~\cite{van2018survey}. PKI cannot mitigate insider attacks; malicious insiders appear benign, already possessing valid keys. Reactive security is application-dependent for e.g.,~state estimation~\cite{pajic2014robustness}, sensor fusion~\cite{hallyburton2023partial}, healthcare~\cite{kwon2014proactive}, and networking~\cite{yeremenko2017secure}. We discuss existing efforts for reactive security in CPS and highlight distinguishing features of our work.

\myparagraph{Distributed data fusion (DDF).}{4pt}{0pt} Trust estimation is distinct from rumor detection in DDF. Rumors arise from networks of connected nodes where information can be double-counted by neglecting to model correlations e.g., unknowingly passing the same data between multiple nodes~\cite{shah2011rumors}. 

\myparagraph{Occupancy grid.}{4pt}{0pt} Multiple agents can use occupancy grids informed by perception data to perform data consistency checking. In~\cite{ambrosin2019design, liu2021miso, zhang2024data}, information is quantized and object existence evaluated on binned binary data.

\myparagraph{Vehicular ad hoc networks (VANETs).}{4pt}{0pt} MBDs detect inconsistencies by correlating data in VANETs. Each agent fuses ownship with nearby agent data. MBDs require that ownship data is absolutely trusted~\cite{van2018survey}; this is only justified if the attacker is only in the network and not on agents. This is too restrictive for real-world cases where no agent is absolutely trusted and adversaries can compromise \emph{any} agent.

Several works on MBDs influenced MATE, including:
\vspace{-4pt}
\begin{itemize}
    \setlength\itemsep{-4pt}
    \item \cite{wang2006autonomous} estimates Beta distributions of trust with binary inputs. Estimation requires trust to be fixed instead of allowing attacks to happen at any time.
    \item \cite{golle2004detecting} detects spoofed poses finding the minimal set of colluding adversaries needed to explain the data. It does not handle natural sensor/algorithm errors.
    \item \cite{bissmeyer2012assessment} maintains trust ``opinions'' of value and confidence tracking visible area with a particle filter using field of view to check for inconsistent observations.
\end{itemize}
\vspace{-4pt}
A shortcoming of MBDs is the requirement that the ego agent is absolutely trusted, making them inapplicable when any agent is vulnerable~\cite{van2018survey}; our security-aware fusion instead estimates trust of \emph{all} agents. A shortcoming of Byzantine models~\cite{theodorakopoulos2004trust,wang2006autonomous} is their inability to handle noisy, error-prone sensor data by requiring unrealistic perfect perception. We solve this challenge with longitudinal filtering.

\myparagraph{Classical data fusion integrity.}{4pt}{0pt} MTT confirms tracks with likelihood ratio tests via recursive track scoring~\cite{1986blackmanRadar}. While scoring improves robustness to natural errors, it provides minimal security and introduces vulnerabilities as proved in~\cite{hallyburton2024bayesian}. Existing approaches to secure fusion cannot handle practical cases with time-varying trust, data uncertainty, prior information, and dynamic field of view. Our approach overcomes these limitations with robust trust estimation.

\section{Collaborative Sensor Fusion} \label{sec:sensor-fusion}

Occlusions due to infrastructure and objects, as in Fig.~\ref{fig:fusion-occluded-views}, challenge SA in safety-critical urban environments. Collaboration between agents in \emph{smart cities} can enhance SA. We consider that mobile ground and static infrastructure agents at urban intersections track objects and send to a data aggregator (AGG) that runs multiple-target tracking (MTT). We describe platform-level algorithms for agent's local inference and discuss centralized MTT for collaborative fusion.

\renewcommand{\subfigwidth}{0.85}
\begin{figure}[!t]
    \centering
    \begin{subfigure}[b]{\linewidth}
        \centering
        \includegraphics[width=0.7\linewidth]{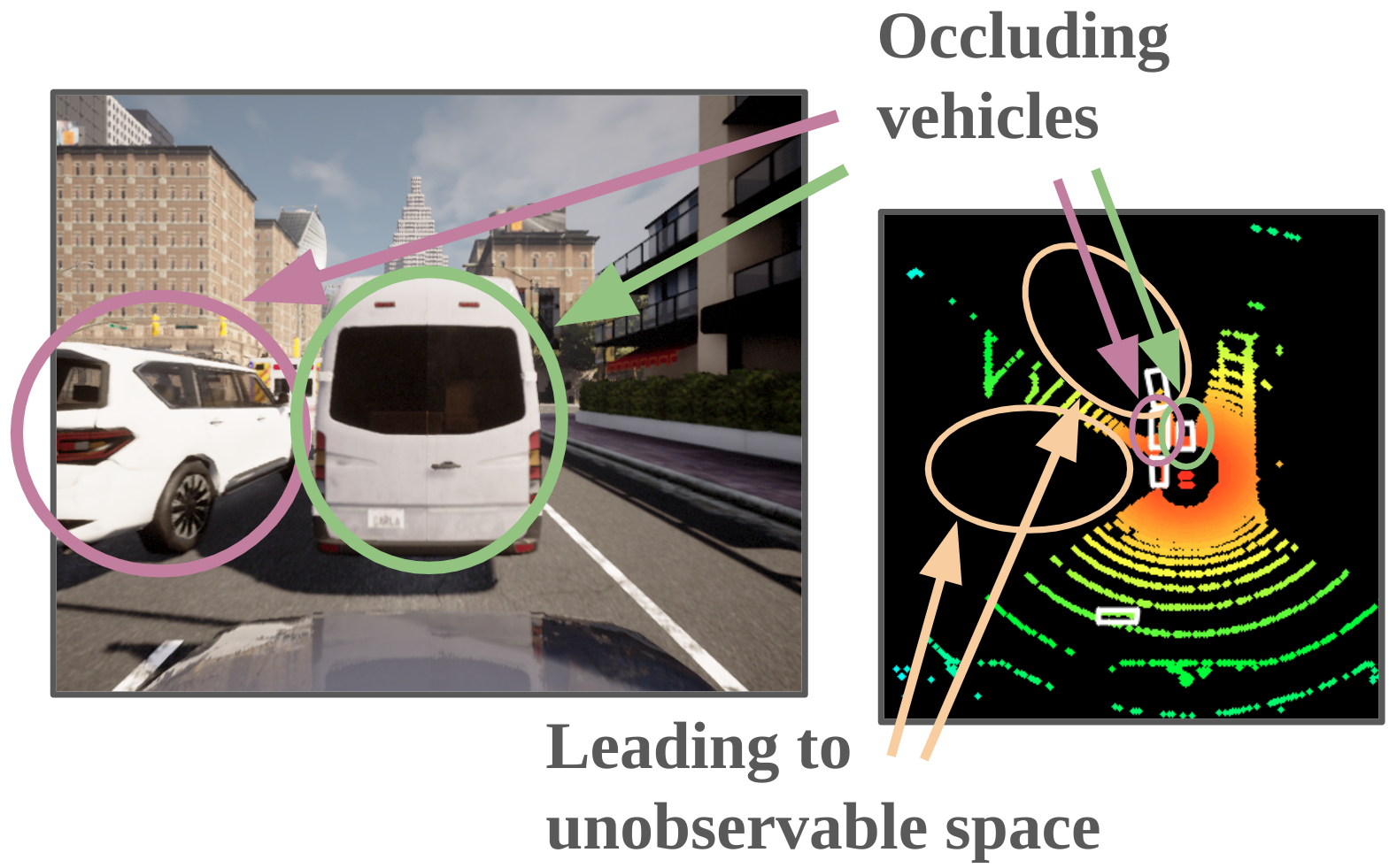}
        \caption{Ego views occluded by other objects.}
    \end{subfigure}
    %
    \begin{subfigure}[b]{\linewidth}
        \centering
        \includegraphics[width=\subfigwidth\linewidth]{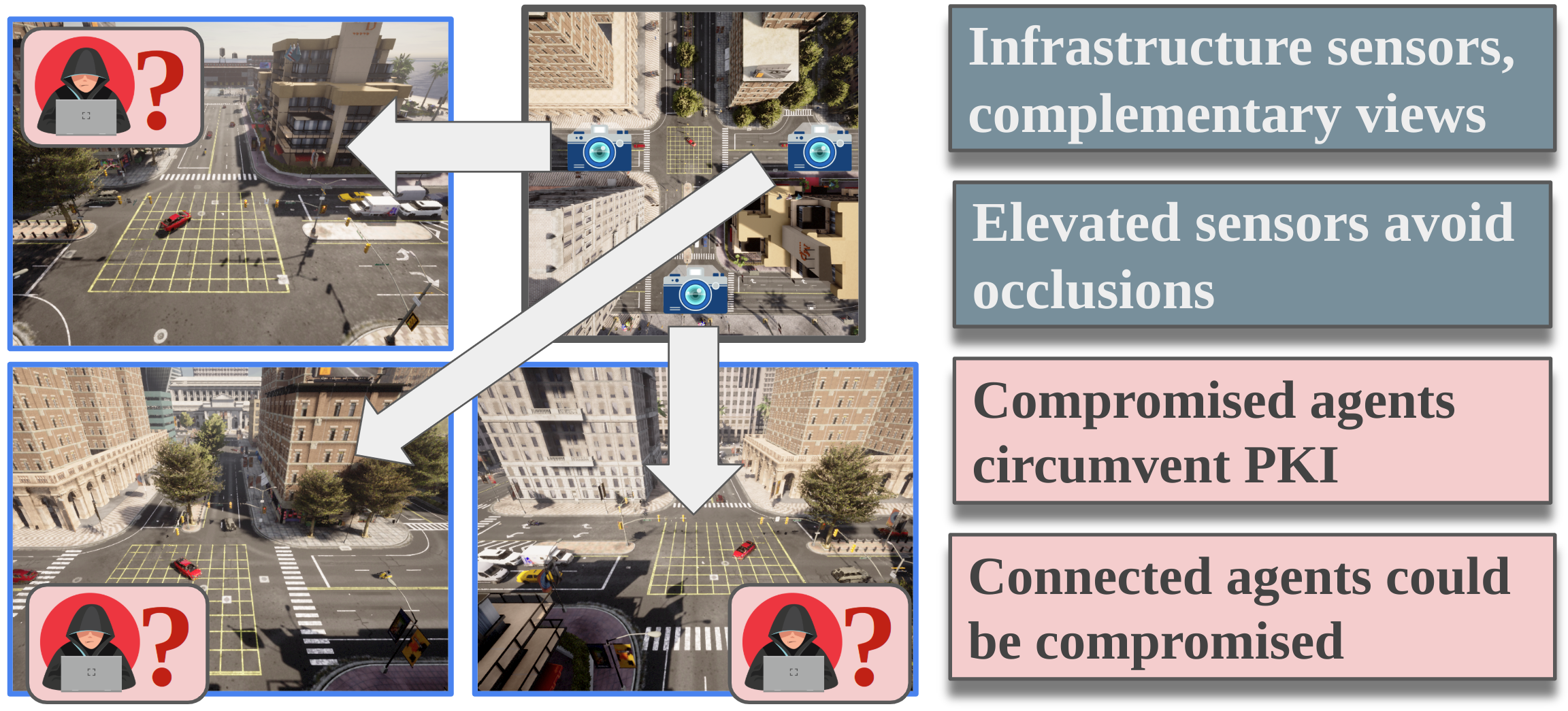}
        \caption{Infrastructure sensors yield complementary views.}
    \end{subfigure}
    \caption{Inter-object occlusions are challenging for AVs and restrict situational awareness (SA), limiting their ability to safely traverse dynamic environments. Multi-agent fusion improves SA but can be adversarially compromised.}
    \label{fig:fusion-occluded-views}
\end{figure}

\subsection{Agent-Local Algorithms}

Agent-local perception detects and local tracking filters objects to mitigate propagation of natural errors to AGG. Filtering with statistical models builds resilience to natural FPs/FNs from noisy sensors. Agents also maintain FOV estimates that dynamically predict sensor visibility. FOV estimates inform AGG of which objects an agent was \emph{expected} to see so as not to penalize for expected missed detections. Algorithms are summarized here with additional details in Appendix~\ref{appendix:fusion}.

\myparagraph{Localization.}{2pt}{0pt} Agents estimate ownship position and attitude with IMU and GPS sensors. Localization pose estimates are tracked in Kalman filter estimators and output at high rate~\cite{bar2004estimation}. 

\myparagraph{Perception.}{2pt}{0pt} Agents detect 3D bounding boxes around objects from LiDAR data. We deploy PointPillars~\cite{lang2019pointpillars} trained on CARLA datasets from~\cite{hallyburton2023datasets}. Ground-based agents and elevated infrastructure agents (Fig.~\ref{fig:fusion-occluded-views}) use different model weights due to differences in scene geometry.

\myparagraph{Tracking.}{2pt}{0pt} An agent's detections feed a local Kalman filter tracker that estimates object states and filters natural FPs/FNs. Detections update position, velocity, orientation, and box size of existing tracks or start new tracks. The tracker model is standard~\cite{1986blackmanRadar} and an implementation is used from AVstack~\cite{hallyburton2023avstack}.

\myparagraph{Field of View.}{2pt}{0pt} Agents use their LiDAR data to estimate (occluded) sensor FOV with algorithms based on ray tracing~\cite{cook1984distributed} and concave hulls~\cite{park2012newconcave}. Implementations are in Appendix~\ref{appendix:fusion}. Fig.~\ref{fig:fov-model} illustrates ray tracing on occluded LiDAR data to form a parametric FOV model in the bird's eye view (BEV).

\begin{figure}[t]
    \centering
    \begin{subfigure}[b]{0.45\linewidth}
        \frame{\includegraphics[width=\linewidth]{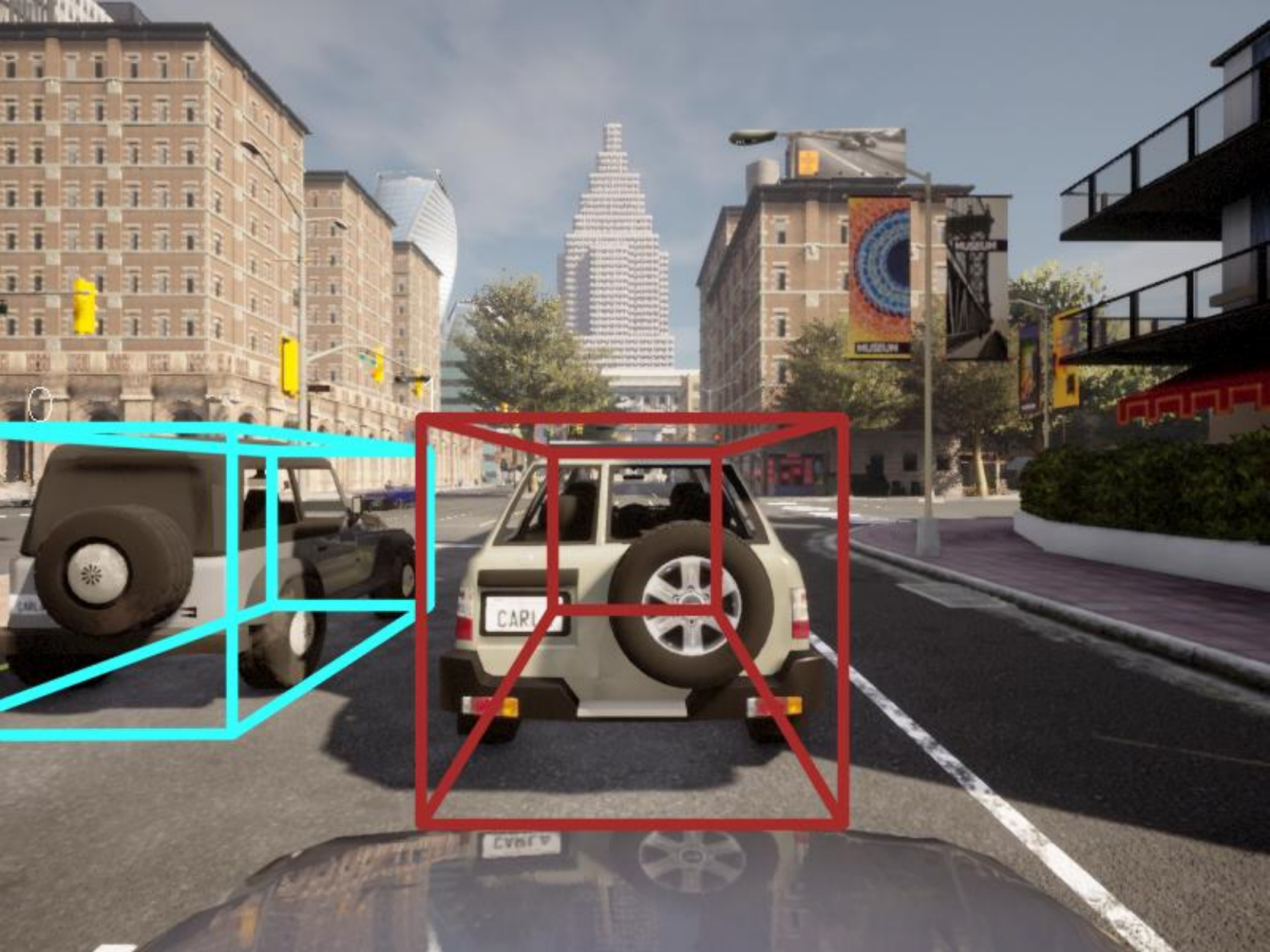}}
        \caption{Front view image}
        \label{fig:fov-model-a}
    \end{subfigure}
    \begin{subfigure}[b]{0.45\linewidth}
        \includegraphics[width=\linewidth,height=3cm]{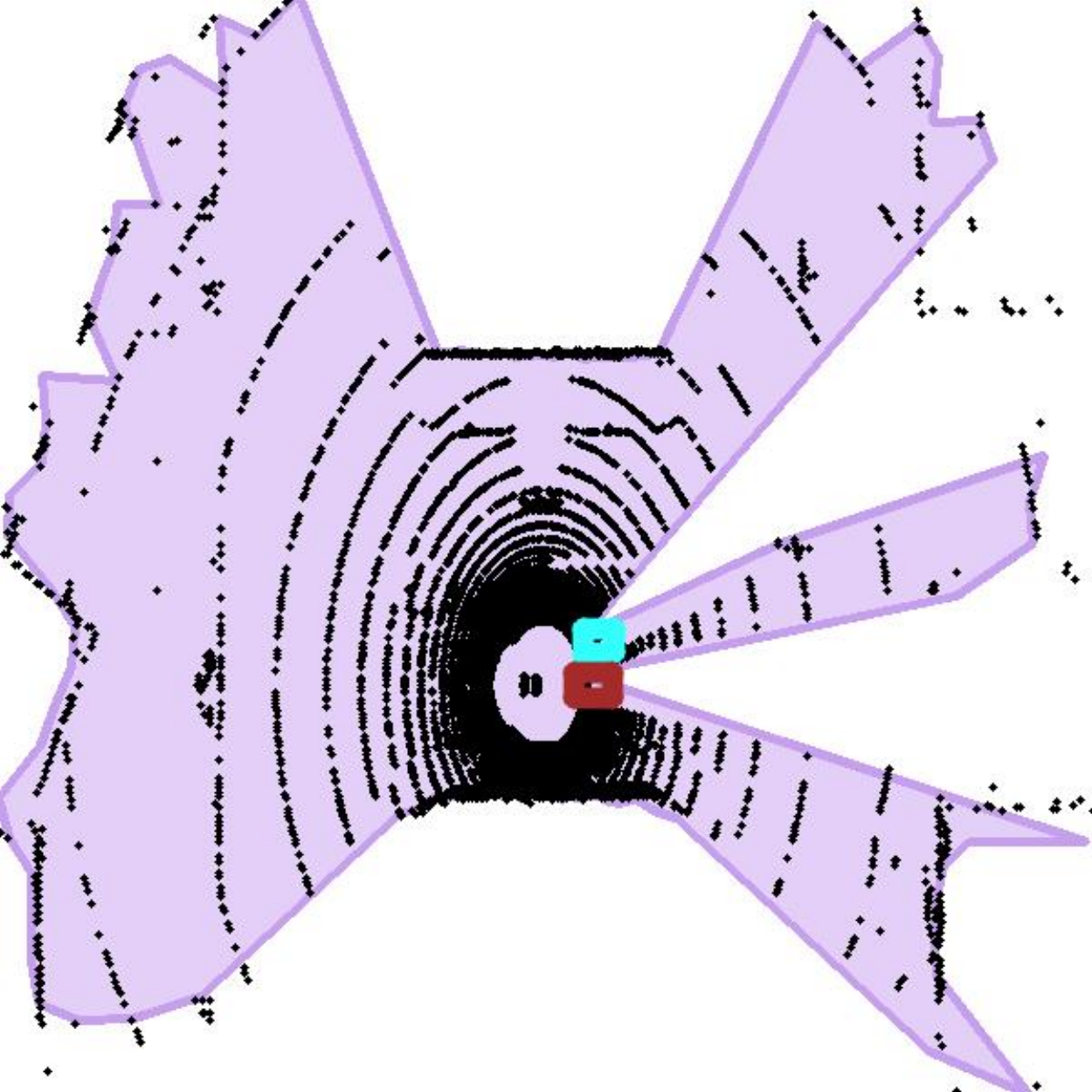}
        \caption{LiDAR point cloud, FOV}
        \label{fig:fov-model-b}
    \end{subfigure}
    \caption{(left) The ego suffers strong occlusion from vehicles and infrastructure. (right) Ray tracing estimates LiDAR FOV with occlusions. The predicted FOV is accurate and enables prediction of (in)visible regions.}
    \label{fig:fov-model}
\end{figure}

\subsection{Multi-Agent, Multiple Target Tracking}

\begin{figure}[t]
    \centering
    \includegraphics[width=0.85\linewidth,trim={2cm 5cm 6cm 5cm},clip,fbox]{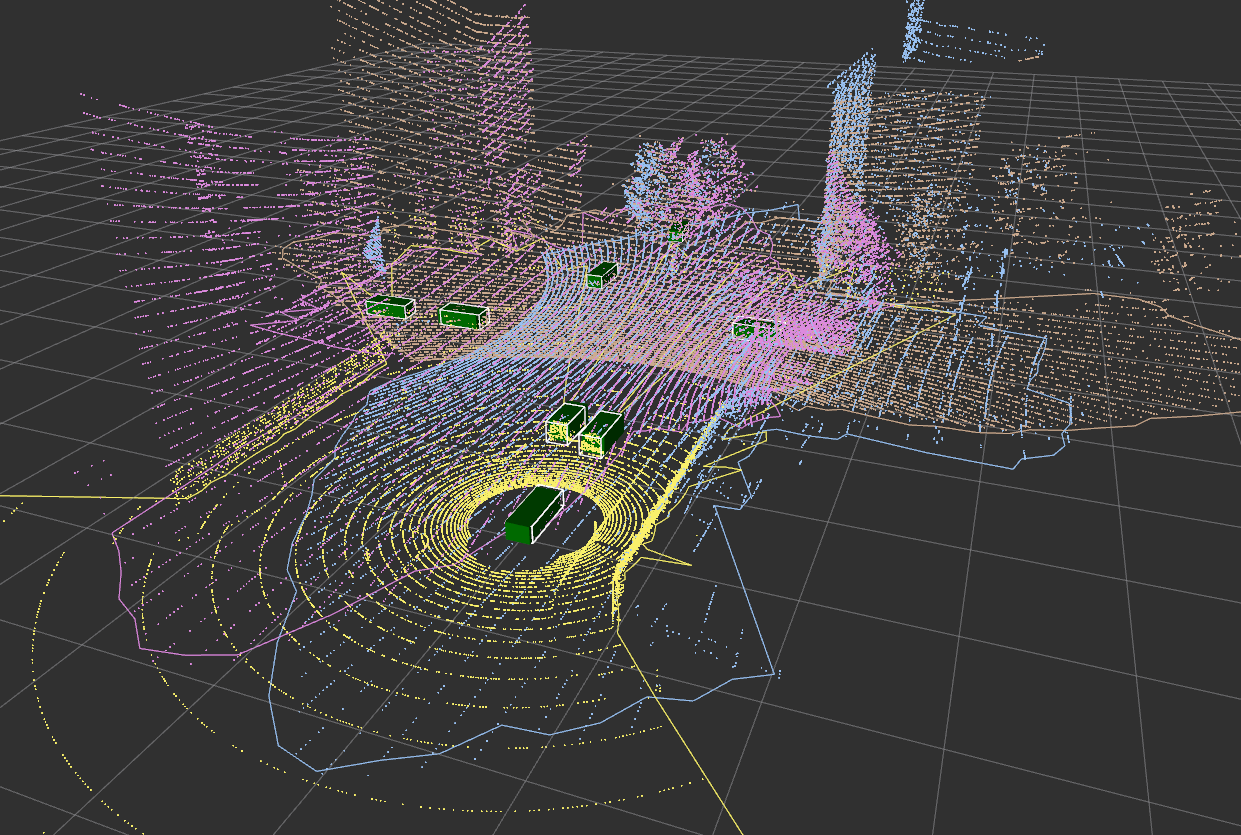}
    \caption{An autonomous agent negotiates a busy urban intersection with the aid of three infrastructure sensors. Colors correspond to data from each agent. Green boxes indicate tracks fused at the data aggregator. Data from all case studies are shown in Figs.~\ref{fig:results-viz-case-0},~\ref{fig:results-viz-case-1}, and~\ref{fig:results-viz-case-2}.}
    \label{fig:multi-agent-fusion-example}
\end{figure}

Agents pass ownship pose, tracked objects, and FOV estimates to trusted computing centers (AGG) that subsequently execute MTT. AGG performs the following: (1) aligning tracked objects from spatially-separated agents by compensating each track pose with agent pose, (2) associating tracked objects from sets of compensated agent tracks using assignment algorithms and affinity metrics (e.g.,~bounding box overlap), and (3) fusing associated data into a global operating picture. Such a pipeline is standard in multi-agent MTT~\cite{1986blackmanRadar,bar1995multitarget,bar2004estimation}. An example outcome of multi-agent MTT is provided in Fig.~\ref{fig:multi-agent-fusion-example}. Four agents represented by unique colors detect objects (box outlines) from LiDAR data (points) while AGG fuses data to get a global operating picture (green filled boxes).

Formally, the multi-agent MTT problem (MA-MTT) is to estimate the object state posterior
\begin{equation} \label{eq:fusion-posterior}
\begin{aligned}
    \Pr(X_{t} | Z_{1:t}, A_{1:t})
\end{aligned}
\end{equation}
where $X_{t}$ are object states for all $i=1...N$ objects, $Z_{1:t}$ are sensor data from all $k=1...K$ agents, and $A_{1:t}$ are characteristics including pose and sensor models. 
\section{Threat Model} \label{sec:threat-model}

At odds with effective multi-agent collaborative fusion are threats against sensing, communication, and computation in multi-agent networks. Security is of particular concern in smart cities and dense urban areas due to the safety-critical nature of high-risk passenger transportation. If attackers execute over physical sensing channels~\cite{2019cao-spoofing,2022hally-frustum}, manipulate agents as insiders~\cite{monteuuis2018my}, or compromise communication~\cite{ansari2021v2x}, they will circumvent proactive security measures including~PKI.

In collaborative data fusion, each agent tracks detections over longitudinal time horizons using spatio-temporal dynamical models to estimate kinematic states. AGG then fuses tracks from all agents with association and filtering algorithms. As such, we categorize threats against agent-level data fusion by two attributes: (1) how the threat manifests in a single frame (``manifest'' characteristic), (2) how the threat evolves over time (``temporal'' characteristic). Evaluating both characteristics is essential for designing security-aware fusion.

\myparagraph{Manifest Characteristic.}{4pt}{0pt} Options for attack manifest at perception include \emph{false positive} (FP), \emph{false negative} (FN), 
and \emph{translated} detections, and at pose, \emph{agent pose errors}. In this work, we consider attacks on perception in both singular (e.g.,~one FP) and multitudinous (e.g.,~many FPs) realizations.

\vspace{-4pt}
\begin{itemize}[leftmargin=12pt]
    \setlength\itemsep{-2pt}
    \item \textbf{False positive:} Fake object(s) injected into detections.
    \item \textbf{False negative:} True object(s) removed from detections.
    \item \textbf{Translation:} Detections are translated spatially.
    \item \textbf{Pose Error:} Agent sends compromised ownship pose.
\end{itemize}
\vspace{-4pt}

\myparagraph{Temporal Characteristic.}{4pt}{0pt} Security analysis of multi-agent data fusion requires evaluation of attack evolution. This is in contrast to perception-level analysis requiring only single-frame examples. We consider temporal evolution of attacks including \emph{static} instances that persist in the same location, \emph{Markovian} instances that evolve with random-walk dynamics, and \emph{trajectory} instances that exhibit predefined trajectories following realistic dynamical models. Attacks not obeying plausible dynamics will be filtered by agent-local algorithms; thus, the attacker must inject longitudinally-consistent traces. 

\vspace{-4pt}
\begin{itemize}[leftmargin=12pt]
    \setlength\itemsep{-2pt}
    \item \textbf{Static:} Threat temporally-invariant, i.e.,~spatially-fixed.
    \item \textbf{Markovian:} Threat evolves with random walk dynamics.
    \item \textbf{Trajectory:} Threat demonstrates a predefined trajectory.
\end{itemize}
\vspace{-4pt}
\section{Security-Aware Sensor Fusion} \label{sec:security-aware}

We discuss the vulnerability of data fusion to motivate a novel security-aware approach. We then define \emph{trustedness} in autonomy and formally state the security-aware sensor fusion problem before providing a summary of our approach.

\subsection{Vulnerability of Classical Data Fusion}

Perception and tracking have been demonstrated vulnerable to attacks~\cite{2019cao-spoofing,2022hally-frustum,jia2020fooling,thys2019fooling,eykholt2018robust}, and in response, agent-level defenses including adversarial training~\cite{sun2020towards}, monocular 3D detection~\cite{hallyburton2023partial}, and inter-sensor correspondence~\cite{liu2021seeing} have been proposed. Unfortunately, adaptive attacks will circumvent platform-level defenses~\cite{athalye2018obfuscated,tramer2020adaptive}. Thus, it is not safe to assume agent-level security can prevent disruption of multi-agent fusion. Moreover, in heterogeneous networks, there is no guarantee each agent is running secure algorithms.

In VANETs, MBDs were proposed to mitigate attacks on collaborative fusion. Unfortunately, MBDs fail to handle natural perception errors~\cite{wang2006autonomous}, require that the agent is absolutely trusted~\cite{van2018survey}, and cannot always recover trusted SA~\cite{golle2004detecting}. Securing multi-agent fusion is still an open problem; we propose an approach based on estimating agent \emph{trustedness}.

\subsection{Trust in Autonomy} \label{sec:security-aware-trust}

To overcome the discussed security challenges in multi-agent SA,
we propose a novel trust-based approach to sensor fusion. Here, we define \emph{trust} and describe the high-level approach before discussing implementation in Secs.~\ref{sec:mate} and~\ref{sec:trusted-fusion}. 

\subsubsection{Defining Trust} \label{sec:security-aware-trust-definition}

We consider the problem of estimating the \emph{trust} of agents and tracks in a network of agents. We use \emph{trust} to represent a mixture of the competency and integrity of data~\cite{huhns2002trusted}. On opposite ends of the spectrum are \emph{trusted} and \emph{distrusted}. An \emph{untrusted} entity is neither trusted nor distrusted. 

\begin{definition}{Trust}
    reflects that provided information from an agent or about an object is (in)consistent with physical laws and the true state of the environment.
\end{definition}

Monitoring trust on \emph{agents} aids in identifying \emph{sources} of misinformation. Monitoring trust on \emph{tracked objects} mitigates ill-advised \emph{actions} downstream caused by the spread of misinformation. Trust can also aid in directing available resources.

\begin{definition}{Trust estimation}
    maps sensor data and agent characteristics to the domain $[0,1]$ where 0 represents complete distrust and 1 complete trust.
\end{definition}

\begin{subdefinition}{Agent trust, $\{\tau^a_k\} \coloneqq \Tau^a$ for agents $k=1...K$.}
    The degree that information contributed by an agent is consistent with trusted agents and physical laws.
\end{subdefinition}

\begin{subdefinition}{Track trust, $\{\tau^c_j\} \coloneqq \Tau^c$ for tracks $j=1...J$.}
    The degree that a track is believed to exist and the estimated state is representative of the true object state.
\end{subdefinition}

\subsubsection{Trust as Bayesian Estimation} \label{sec:security-aware-trust-bayesian}

We pose a joint trust estimation \& sensor fusion problem building on the posterior from Eq.~\eqref{eq:fusion-posterior}. Adding trust results in $\Pr(X_{t}, \Tau^c_{t}, \Tau^a_{t} | Z_{1:t}, A_{1:t})$. Conditional probability decomposes the posterior into two subproblems
\begin{equation}\label{eq:trust-posterior}
\begin{aligned}
    \Pr(&X_{t}, \Tau^c_{t}, \Tau^a_{t} | Z_{1:t}, A_{1:t}) \\
     &= \Pr(\Tau^c_{t}, \Tau^a_{t} | Z_{1:t}, A_{1:t}) \Pr(X_{t} | \Tau^c_{t}, \Tau^a_{t}, Z_{1:t}, A_{1:t}). 
\end{aligned}
\end{equation}
The joint problem can now be solved via \textbf{(1) trust estimation} (Sec.~\ref{sec:mate}) and \textbf{(2) trust-informed data fusion} (Sec.~\ref{sec:trusted-fusion}). 

\subsection{Overview of Security-Aware Fusion} \label{sec:security-aware-overview}

\begin{figure}[t]
    \centering
    \includegraphics[width=0.94\linewidth]{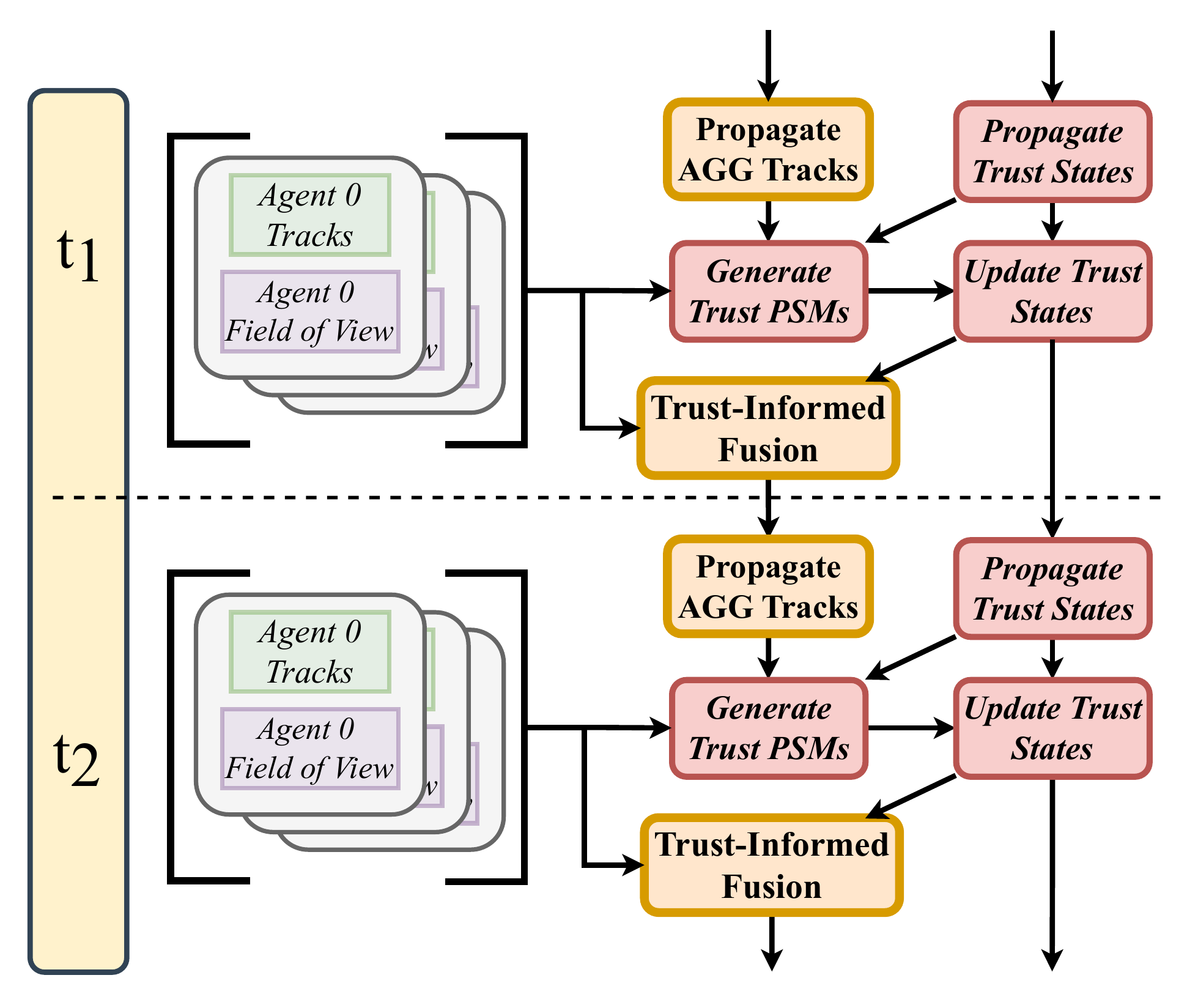}
    \caption{Trust estimation and trust-informed fusion operate sequentially within a security-aware fusion architecture.}
    \label{fig:trust-full-logic}
\end{figure}

\begin{figure*}[t]
    \centering
    \includegraphics[width=0.9\linewidth]{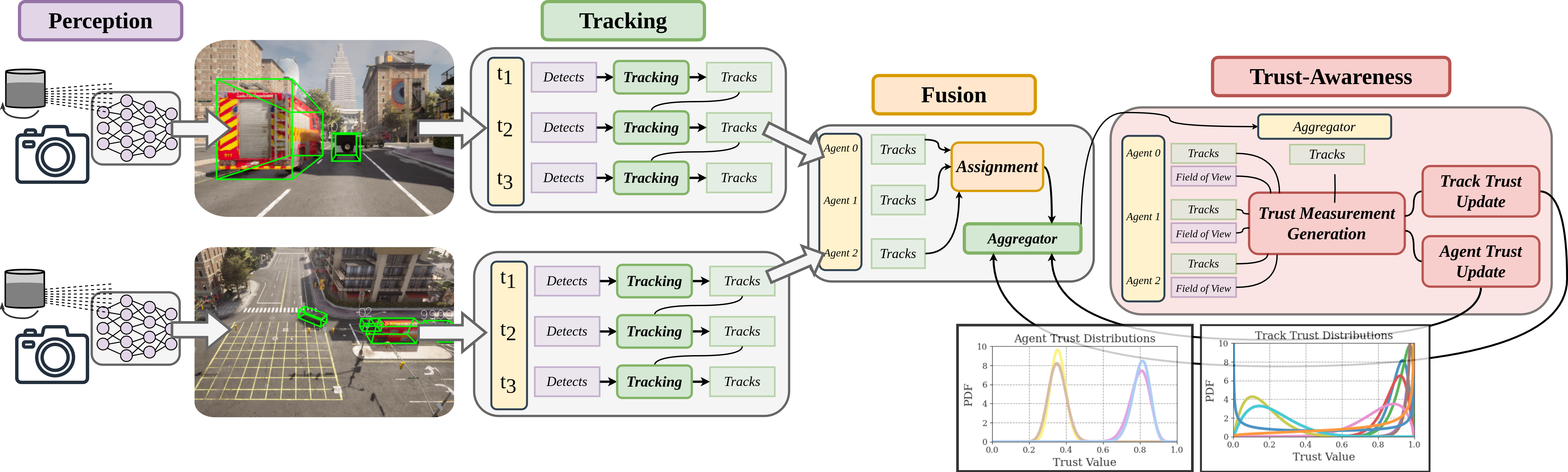}
    \caption{Traditional sensor fusion is vulnerable to attacks on small numbers of agents even in a large multi-agent network. We introduce \emph{trust awareness} into the fusion pipeline to detect, identify, mitigate, and recover from attacks to secure safety-critical infrastructure. Bayesian methods for trust awareness output distributions over trustedness for agents and tracked objects.}
    \label{fig:fusion-trust-aware}
\end{figure*}

Fig.~\ref{fig:trust-full-logic} summarizes our security-aware sensor fusion (as a time-series representation), while Fig.~\ref{fig:fusion-trust-aware} illustrates the flow of information. The agent-local pipelines are undisturbed from their implementations presented in Sec.~\ref{sec:sensor-fusion}; this importantly relieves needing to modify agent source code which can be challenging in heterogeneous networks. Security-aware fusion introduces two new elements at AGG: (1)~trust estimation, and (2)~trust-informed data fusion.


\myparagraph{Trust estimation.}{2pt}{0pt} Trust estimates for agents and tracks are maintained as PDFs on $[0,1]$. Tracks, poses, and FOVs from agents and SA from AGG are mapped to \emph{pseudomeasurements} (PSMs) of trust that update distributions using an HMM via propagation and update~steps. Sec.~\ref{sec:mate} describes our novel approach: MATE, the 
multi-agent trust estimator.

\myparagraph{Trust-informed fusion.}{2pt}{0pt} With trust PDFs in hand, AGG performs trust-informed data fusion to update global SA. Each agent's influence is weighted by its trust PDFs ensuring distrusted agents cannot significantly alter SA. Track trust PDFs influence downstream decision logic by flagging or removing distrusted tracks. Sec.~\ref{sec:trusted-fusion} describes trust-informed fusion.
\section{Multi-Agent Trust Estimation} \label{sec:mate}

We present MATE, the multi-agent trust estimator, designed to solve subproblem (1) of Eq.~\eqref{eq:trust-posterior} by estimating agent and track trust PDFs. In a Bayesian context, this amounts to solving
\begin{equation} \label{eq:trust-estimation-posterior}
    \begin{aligned}
        \Pr(\Tau^c_{t}, \Tau^a_{t} & | Z_{1:t}, A_{1:t}) \propto \\
        & \Pr(Z_{1:t}, A_{1:t} | \Tau^c_{t}, \Tau^a_{t}) \Pr(\Tau^c_{t} | \Tau^a_{t}) \Pr(\Tau^a_{t}).
    \end{aligned}
\end{equation}
MATE jointly estimates agent and track trust by specifying prior density over trust $\Pr(\Tau^c_{t} | \Tau^a_{t}),\, \Pr(\Tau^a_{t})$ and data likelihood functions $\Pr(Z_{1:t}, A_{1:t} | \Tau^c_{t}, \Tau^a_{t})$ to form the trust posterior. MATE operates in the context of an HMM with four stages: (0) initializing trust PDFs via prior information, (1)~temporal propagation of the most recent trust PDFs, (2)~generating PSMs of trust, and (3)~trust PDF updates from~PSMs.

\subsection{Step 0: Initial Priors on Trust} \label{sec:mate-priors}

An attractive facet of Bayesian estimation is the natural incorporation of prior information. Such an advantage is unavailable in many MBD works. The two-parameter Beta distribution is a natural choice for a prior, as first proposed in~\cite{hallyburton2024bayesian}, as it is defined on the trust domain $[0,1]$ and flexibly has uniform, Gaussian, and exponential tendencies. Moreover, it supports conjugacy conditions discussed in Section~\ref{sec:mate-update}.

Priors can reduce the minimum number of uncompromised agents needed to obtain trusted SA. Previous works were limited to no more than half of agents compromised (e.g.,~\cite{pajic2014robustness}). Strategically, highly trusted agents can be deployed to outweigh agents of unknown trust. Similarly, giving new agents high-variance priors will mitigate their initial impact on data fusion before a reliable trust estimate is formed.

\subsection{Step 1: Trust State Propagation} \label{sec:mate-propagation}

A transition model propagates trust PDFs from the previous to the current time reflecting that trust is a dynamic and evolving state. To our knowledge, no transition models of multi-agent trust exist, so we build three transition models. Models follow the intuition that, in the absence of measurements, the uncertainty on an entity's trust should increase over time. For a Beta distribution, we use the following notation: $\text{Beta}=f_B(\alpha,\beta)$, expectation $\expectation[f_B] \coloneqq \mu = \frac{\alpha}{\alpha + \beta}$, precision $\precision[f_B]\coloneqq \nu = \alpha + \beta$. 

\myparagraph{Prior interpolation propagator (PropP).}{4pt}{0pt} Over time, PropP gradually returns the trust to the initial prior. PropP takes parametric distributions for current trust and the initial prior trust and incrementally interpolates their parameters with a tuning parameter, $\omega$. In particular, for the Beta, this is
\begin{equation} \label{eq:trust-propagator-prior}
    \begin{aligned}
        \alpha_{t} = (1-\omega) \, \alpha_{t-1} + \omega \, \propagatorprioralpha, ~~~
        \beta_{t} = (1-\omega) \, \beta_{t-1} + \omega \, \propagatorpriorbeta
    \end{aligned}
\end{equation}

\myparagraph{Expectation propagator (PropE).}{4pt}{0pt} PropE updates $\expectation[f_B]$ with a predefined change factor $\Delta \mu$ towards some target $\Bar{\mu}$ without changing $\precision[f_B]$. In particular, $\mu_{t}=\mu_{t-1} + \sfrac{1}{\Delta \mu}(\Bar{\mu} - \mu_{t-1})$ with bounds enforced as $0 \leq \mu \leq 1$ and $\Bar{\mu}=0.5$ as the target. In a Beta distribution, the parameters are then set as $\alpha_{t} = \mu_{t} \nu_{t-1}$ and $\beta_{t} = (1-\mu_{t}) \nu_{t-1}$ in the standard way.

\myparagraph{Variance propagator (PropV).}{4pt}{0pt} PropV updates $\precision[f_B]$ towards target $\Bar{\nu}$ at constant $\expectation[f_B]$ in the same way as PropE.

Propagation models can be composed to achieve, for example, simultaneous expectation and variance propagation. Including trust state propagation is important to account for the dynamic nature of a trust state and that agents/tracks become less trustworthy in the absence of measurements.

\subsection{Step 2: Trust Pseudomeasurements (PSMs)} \label{sec:mate-psm}

Bayesian trust estimation requires a likelihood function, $\Pr(Z_{1:t},A_{1:t}|\Tau^a_{t},\Tau^c_{t})$ (Eq.~\eqref{eq:trust-estimation-posterior}). There is no singular likelihood due to both the inexactness of defining trust and that there is no ``trust sensor'' providing trust measurements. Instead, we define our own \emph{trust PSM functions} that map perception-oriented sensor data to trust measurements.

\begin{definition}{Trust pseudomeasurement (PSM), $\rho^c_j,\,\rho^a_k$ for track $j$, agent $k$, implicitly at time $t$.}
    A PSM is a measurement of trust as if from a trust sensor and is a function of sensor data, agent characteristics, and estimated trust PDFs.
\end{definition}

\subsubsection{PSM as a (Value, Uncertainty) Pair}

Trust PSMs are composed of a value and uncertainty to emulate the output of a sensor (e.g.,~GPS gives position measurement and uncertainty). We notate track PSMs as $\rho^c_j=(v^c_{j}, c^c_{j})$ where $v^c_{j}\in [0, 1]$ is the ``value'', and $c^c_{j}\in [0,1]$ is the ``confidence'' (inverse uncertainty) and agent PSMs as $\rho^a_k=(v^a_{k}, c^a_{k})$.

\subsubsection{Building PSM Functions}

There is no singular definition of trust and thus no singular implementation of PSM functions. We build functions to match intuition that trust is based on data agreement with trusted agents. Let $g^c,\, g^a$ denote track/agent-focused PSM functions that map data to the trust domain of $[0,1]$. Specifically,
\begin{equation} \label{eq:trust-psm-function}
\begin{aligned}
    \rho^c_j = g^c_j(\Tau^a_{t}, Z_{1:t}, A_{1:t}), \quad \rho^a_k = g^a_k(\Tau^c_{t}, Z_{1:t}, A_{1:t})
\end{aligned}  
\end{equation}
with $\Rho^c \coloneqq \{\rho^c_j\}_{j=1}^J$ and $\Rho^a \coloneqq \{\rho^a_k\}_{k=1}^K$. Uncompromised agents should consistently track true objects within their FOVs. Conversely, any ``object'' purportedly tracked by some but not all agents is either an FP in the set of agents observing the object or an FN in the set of agents not observing. This philosophy is sufficient to construct our own PSM functions that evaluate agreement between agents and AGG. In what follows, we describe how we build the functions $g^c,\, g^a$. 

\begin{figure}[t]
    \centering
    \includegraphics[trim={0cm 0cm 0cm 0cm},clip,width=0.95\linewidth]{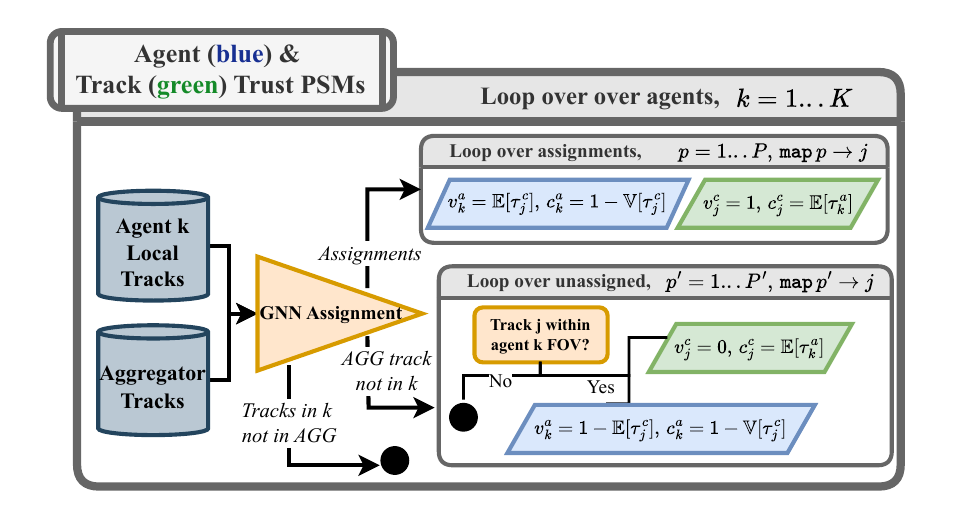}
    \caption{PSMs for tracks/agents leveraging assignment between local agent and aggregated tracks. Consistency/deviation from expected observation models provide affirmative/negative PSMs. Black circles represent no action. PSMs recursively depend on both agent and track trust.}
    \label{fig:psm-logic}
\end{figure}

\begin{figure*}[t]
    \centering
    \includegraphics[trim={0cm 0cm 0cm 0cm},clip,width=0.9\linewidth]{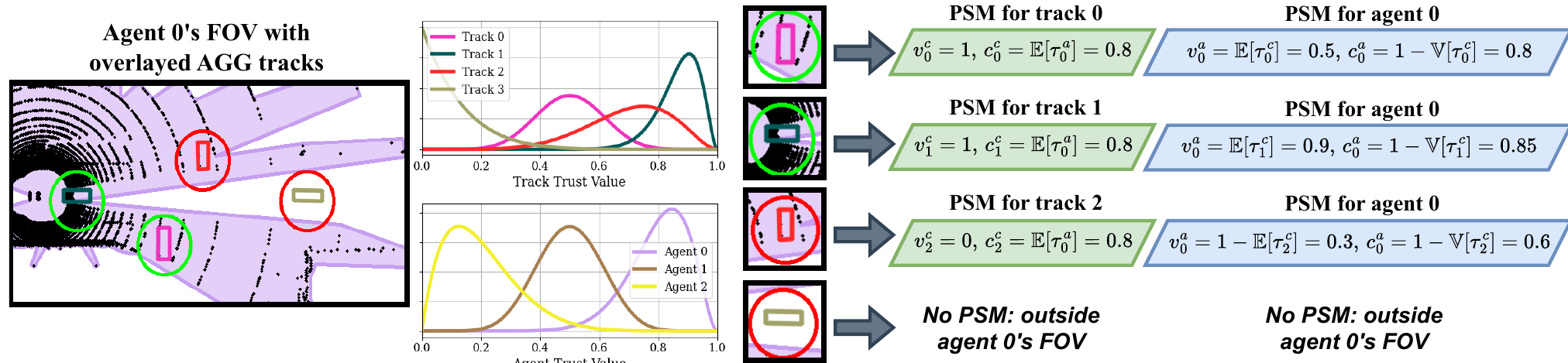}
    \caption{Example PSM generation following algorithm in Fig.~\ref{fig:psm-logic}. Agent 0 (A0) has two detections matching aggregator tracks (T0: pink, T1: green) and misses two AGG tracks (T2: red, T3: yellow). T0 and T1 obtain positive PSMs due to A0 detections with confidence as A0's trust. T2 obtains negative PSM with A0's trust as confidence. A0 receives three PSMs with value deriving from T0 trust, T1 trust, and inverse T2 trust. T3 is outside A0's FOV so does not contribute to PSMs.}
    \label{fig:psm-case}
\end{figure*}

The full PSM logic is in Fig.~\ref{fig:psm-logic}. Agent tracks are matched with AGG tracks via an assignment (e.g., the standard Munkres or JVC algorithms~\cite{crouse2016implementing}). Matches are scored favorably while unassigned AGG tracks may penalize the agent and/or track trust. Specifically, if AGG track $j$ in assignment $p$ is seen by agent $k$, the track receives an affirming PSM ($v^c_{j}=1$) with confidence as the agent's trust ($c^c_{j}=\expectation[\tau^a_k]$). The agent receives a PSM with value of the track trust ($c^a_{k}=\expectation[\tau^c_j]$) and confidence the inverse track's uncertainty ($1-\variance[\tau^c_j])$ with $\variance$ denoting variance. Unmatched AGG tracks, if expected to have been seen by the agent, provide negative PSMs to tracks ($v^c_{j}=0$) and PSMs to agents with value of the inverse track trust ($1-\expectation[\tau^c_j]$). To build an intuition for PSMs, Fig.~\ref{fig:psm-case} walks through a PSM-building case study.

\subsubsection{PSMs Depend on FOV Model}

PSM functions need estimates of each sensor's FOV because they require predicting \emph{what an agent is expected to see}. The FOV varies with time and is influenced by agent pose, sensor characteristics, and occlusions from objects/infrastructure. Prior works on MBDs model FOV as fixed circular shapes (e.g.,~\cite{golle2004detecting,soleymani2017secure,tsukada2022misbehavior,allig2019trustworthiness}). While an unobstructed sensor may have a circular FOV, objects/infrastructure truncate visibility and create unobservable regions of space; a circular model will perform poorly in such contexts and may unfairly penalize trust PDFs. Instead, MATE leverages the dynamic FOV estimates computed at the agent-level as discussed in Sec.~\ref{sec:sensor-fusion}.


\subsection{Step 3: Trust State Updates} \label{sec:mate-update}

Equipped with PSMs derived from sensor data, the final step of MATE is to mix the previous trust state propagated to the current time with the PSMs. In the context of Bayesian parameter estimation, the propagated previous state estimate acts as the current time's ``prior''. The posterior is proportional to the prior and the measurement likelihood. 

\subsubsection{Alternating Updates via Gibbs Decomposition}

We tackle the joint problem of estimating the agent \emph{and} track trust posterior, $\Pr(\Tau^c_{t}, \Tau^a_{t} | Z_{1:t}, A_{1:t})$, with an alternating two-step process inspired by the popular Gibbs sampling (see e.g.,~\cite{casella1992explaining} for an overview of Gibbs). The alternation strategy leverages conditional probabilities, i.e., repeating
\begin{equation}\label{eq:trust-posterior-gibbs}
\begin{aligned}
    (1)\ &\text{Estimate track trust:} \ \Pr(\Tau^c_{t}\ |\ \Tau^a_{t-1}, Z_{1:t}, A_{1:t}) \\
    (2)\ &\text{Estimate agent trust:} \ \Pr(\Tau^a_{t}\ |\ \Tau^c_{t}, Z_{1:t}, A_{1:t}).
\end{aligned}
\end{equation}
It is well known that Gibbs sampling converges to the true posterior when the underlying distribution is stationary~\cite{casella1992explaining}. However, trust distributions may not be stationary because attacks can happen at any time. Thus, we cannot formally prove convergence. However, in experiments, we observe consistent convergence of the model over short time horizons. We present theoretical justification for the Gibbs approach and discuss convergence in Appendix~\ref{appendix:mate-gibbs}.

\subsubsection{Independence Assumption}

The trust posterior is a PDF of agent and track trust on $[0,1]$ with 0 as \emph{distrusted} and 1 as \emph{trusted}. The full distribution falls on a hypercube with domain $[0,1]^{J\times K}$ for $J$ tracks and $K$ agents. Since, as in Eq.~\eqref{eq:trust-posterior-gibbs} and Fig.~\ref{fig:psm-logic}, a trust PSM of an agent leverages the trust state of a track and vice-versa, there may exist correlations between trust states of agents and tracks (but not between kinematic track states from each agent).

Unfortunately, we know of no computationally efficient way to model the full joint distribution since, to our knowledge, there is no appropriate parametric model for multi-dimensional trust. Instead, it comes at great computational convenience to reduce the dimensionality from the $J\times K$ dimensional hypercube to $JK$ one-dimensional estimates to be modeled each by Beta distributions. Approximating that each trust state is independent leads to
\begin{equation}\label{eq:trust-independence}
\begin{aligned}
\Pr(\Tau^c_{t}\ |\ \Tau^a_{t-1}, Z_{1:t}, A_{1:t})&=\Pi_{j=1}^C \Pr(\tau^c_{j,t}\ |\ \Tau^a_{t-1}, Z_{1:t}, A_{1:t})\\
\Pr(\Tau^a_{t}\ |\ \Tau^c_{t}, Z_{1:t}, A_{1:t})&=\Pi_{k=1}^K \Pr(\tau^a_{k,t}\ |\ \Tau^c_{t}, Z_{1:t}, A_{1:t}).
\end{aligned}
\end{equation}
The choice of a Beta prior and Bernoulli likelihood yield a Beta posterior; this has attractive and intuitive properties for trust estimation and importantly a closed-form analytical solution. We discuss the benefits and drawbacks of the independence assumption further in Appendix~\ref{appendix:mate-bayesian}.

\subsubsection{Efficient Updates with Beta-Bernoulli Conjugacy}

PSM functions from Eq.~\eqref{eq:trust-psm-function} allow us to approximate the originally intractable posterior from Eq.~\eqref{eq:trust-estimation-posterior} as
\begin{equation} \label{eq:trust-posterior-with-psms}
\begin{aligned}
\Pr(\Tau^c_{t} | \Tau^a_{t-1}, Z_{1:t}, A_{1:t}) & \approx \Pr(\Tau^c_{t} | \Rho^c_{1:t}) \\
\Pr(\Tau^a_{t} | \Tau^c_{t}, Z_{1:t}, A_{1:t}) & \approx \Pr(\Tau^a_{t} | \Rho^a_{1:t}).
\end{aligned}  
\end{equation}
PSMs can then be used to directly update estimates of the trust posterior. Using a Beta distribution (two parameters, $\left(\alpha_0,\,\beta_0\right)$), for the prior and a Bernoulli likelihood on the PSMs yields a Beta posterior - a ``conjugate pair''. The parameter update for the pair has a closed form; the derivation is in Appendix~\ref{appendix:mate-bayesian}. With PSMs as a tuple, $\rho=(v, c)$, the update to the two-parameter trust PDF of e.g., track trusts will be
\begin{equation}\label{eq:beta-bernoulli-update}
    \begin{aligned}
        \alpha^c_{j,t} &= \alpha^c_{j,t-1} + \Delta \alpha^c_{j,t},\quad \Delta \alpha^c_{j,t} = \sum_i c^c_{j,i} \, v^c_{j,i} \\
        \beta^c_{j,t} &= \beta^c_{j,t-1} + \Delta \beta^c_{j,t},\quad \Delta \beta^c_{j,t} = \sum_i c^c_{j,i} \, (1-v^c_{j,i})
    \end{aligned}
\end{equation}
where the $i$ subscript is over each element in the list of PSMs. The resulting $(\alpha^c_{j,t},\beta^c_{j,t})$ are the new parameters for track $j$'s trust. The same process applies for agent trust, $(\alpha^a_{k,t}, \beta^a_{k,t})$.

\subsubsection{Negatively-Weighted Update.}

The Beta parameters $(\alpha,\,\beta)$ act as the number of positive and negative observations. The update in Eq.~\ref{eq:beta-bernoulli-update} weighs positive and negative PSMs equally. In security, this is undesirable. Instead, trust ought to be conservatively built given only large volumes of consistent data while trust ought to be lost given even small amounts of inconsistent data~\cite{huhns2002trusted}. This motivates a negatively-weighted trust update; e.g., for track trust
\begin{equation}\label{eq:beta-bernoulli-update-weighted}
    \begin{aligned}
        \begin{aligned}
            \Delta \alpha^c_{j,t} = \sum_i c^c_{j,i} &\, v_{j,i}, \quad \Delta \beta^c_{j,t} = \sum_i \omega^c_{j,i} \, c^c_{j,i} \, (1-v^c_{j,i})\\
            \omega^c_{j,i} &= \left.\begin{cases}
                \tracknegativitybias & v^c_{j,i} < \tracknegativitythreshold \\
                1.0 & \text{otherwise} 
            \end{cases} \right\}
        \end{aligned}
    \end{aligned}
\end{equation}
where $\tracknegativitybias$ and $\tracknegativitythreshold$ are the \emph{track negativity bias/threshold}, respectively. Suppose $v^c_{j,i}=0.9$, $\tracknegativitybias=9$, and $\tracknegativitythreshold=1.0$ (i.e.,~no threshold). Such a bias would cause an equality between $\Delta \alpha^c_{j,t}$ and $\Delta \beta^c_{j,t}$ despite that the PSM is undoubtedly positive. Instead, $\tracknegativitythreshold<1$, acts to enhance decidedly negative PSMs.  The agent-oriented weighted update uses the agent negativity bias/threshold pair $(\agentnegativitybias,\,\agentnegativitythreshold)$. While the weighted update no longer strictly obeys the conjugacy condition, it is more closely aligned with the intuition behind trust building.
\section{Trust-Informed Data Fusion} \label{sec:trusted-fusion}

MTT tackles target \emph{existence determination} (i.e.,~real or false tracks) and \emph{state estimation} (e.g.,~position, velocity states) as discussed in Sec.~\ref{sec:sensor-fusion}. The final step of security-aware sensor fusion is to apply estimated trust PDFs from MATE to MTT tasks to form trust-informed data fusion. Relevant to existence determination is \emph{trust-based track thresholding} while \emph{trust-weighted state updates} applies trust to state estimation.

\subsection{Trust-Based Track Thresholding}

If a tracked object is distrusted, it may be an FP or possess inaccurate estimated state. A natural application of trust to existence determination is to single-out tracks of low trust or highly uncertain trust in downstream decision-making. Any track with a low trust (i.e.,~$\expectation[\tau^c_j] < \thresholdtrackignore$) is flagged. It is important to maintain the track's state internally and not delete it (i.e.,~keep it in the AGG but e.g.,~flag it for additional scrutiny) to continue to update agent trusts given all information.

\subsection{Trust-Weighted State Updates}

Interacting agents ought to be significantly \emph{trusted} to gain privileges of contributing to the operating picture for a safety-critical mission~\cite{huhns2002trusted}. In the case of translation attacks where tracks are gradually drifted off over time or of attacks on non-position states (e.g.,~vehicle orientation) that do not affect position-based measurement-to-track assignment, a trust-weighted approach to a track's state update is warranted. This allows all objects to continue to be estimated by trustworthy agents while discounting distrusted agents. 

To perform trust-weighted updates, the Kalman gain matrix for updating estimated object state in MTT ($\mathcal{K}$) is weighted by agent trust, $\expectation[\tau^a_k]$. We add a tuning parameter, $\weightedkalmanupdate$, so that the weighted gain is $\expectation[\tau^a_k]^{\weightedkalmanupdate} \, \mathcal{K}$. Since $\expectation[\tau^a_k]\in[0,1]$, the weight is pushed towards 1 when $\weightedkalmanupdate<1$ and towards 0 when $\weightedkalmanupdate>1$.

\vspace{4pt}
With (1) MATE to specify prior trust, propagate trust estimates, derive PSMs from sensor data, and update trust estimates, and (2) trust-informed data fusion for trust-based track thresholding and trust-weighted state updates, we fully specify the security-aware sensor fusion pipeline.
\section{Experiments in Unreal Engine Smart Cities} \label{sec:exp-ue}

We test security-aware fusion in the physics-based Unreal Engine simulator, CARLA. We construct safety-critical scenes where a mixture of static and mobile agents negotiate smart city intersections with occlusions and adversaries. Starting from baseline benign (unattacked) scenes, we generate adversarial versions through Monte Carlo (MC) adversary simulation, perform trust algorithm tuning on MC training datasets, and run trust case studies on testing scenes. We test both baseline \emph{and} attack scenes both with (``Security-Aware'') \emph{and} without (``No Security'') our security aware algorithms. 

\subsection{Multi-Agent (Adversarial) Datasets} \label{sec:exp-ue-dataset}

Security-aware fusion functions when agents have (partially) overlapping FOVs. We generate longitudinal datasets of mixtures of diverse agents negotiating smart-city intersections. We use the recently released framework for creating multi-sensor, multi-agent datasets from~\cite{hallyburton2023datasets} and apply adversary models in postprocessing. Fig.~\ref{fig:case-diagrams} provides example scene visualizations while Table~\ref{tab:dataset-cases} in Appendix~\ref{appendix:datasets} describes unique scene characteristics. An example attack trace applied to an unattacked scene is in Fig.~\ref{fig:multi-agent-ros-cc}. 3D projections of the scenes accompany the trust results in Figs.~\ref{fig:results-viz-case-0},~\ref{fig:results-viz-case-1}. 
and~\ref{fig:results-viz-case-2}.

\renewcommand{\subfigwidth}{0.45}
\begin{figure}[t]
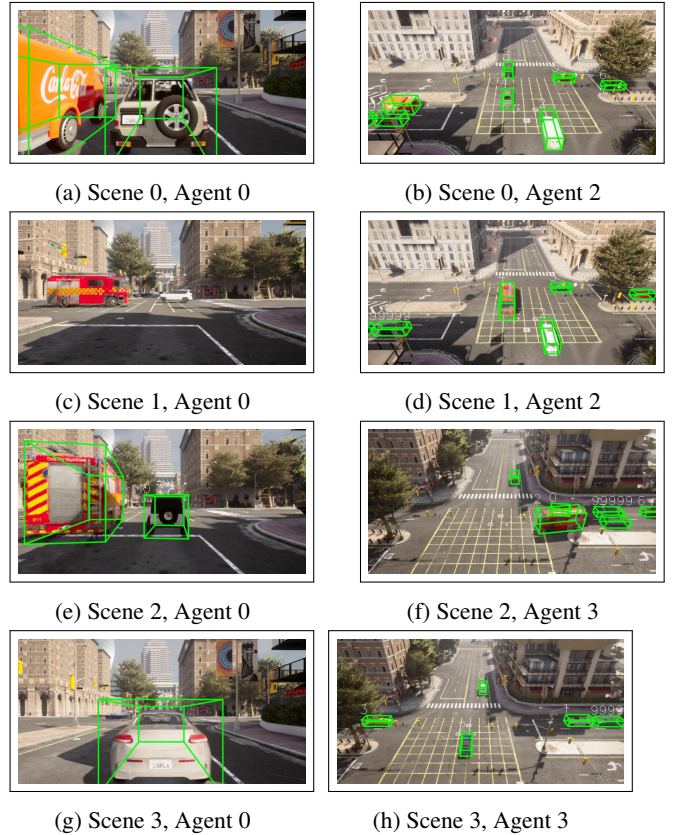

    \centering
    \foreach \scene in {0,1}
    {
        \foreach \agent in {0,2}
        {
            \begin{subfigure}[b]{\subfigwidth\linewidth}
                \centering
                \includegraphics[width=\linewidth,trim={2cm 4.5cm 2cm 4.5cm},clip,fbox]{diagrams/case_study_visualizations/scene\scene/agent\agent_frame302_image.pdf}
                \caption{Scene \scene, Agent \agent}
            \end{subfigure}
            \hfill
        }
    }
    \foreach \scene in {2,3}
    {
        \foreach \agent in {0,3}
        {
            \begin{subfigure}[b]{\subfigwidth\linewidth}
                \centering
                \includegraphics[width=\linewidth,trim={2cm 4.5cm 2cm 4.5cm},clip,fbox]{diagrams/case_study_visualizations/scene\scene/agent\agent_frame302_image.pdf}
                \caption{Scene \scene, Agent \agent}
            \end{subfigure}
            \hfill
        }
    }
    \caption{Case studies each use four agents with complementary vantage points on the scene. Each scene is a unique realization of a safety-critical smart-city intersection.}
    \label{fig:case-diagrams}
\end{figure}

\renewcommand{\subfigwidth}{0.64}
\begin{figure}[t]
    \centering
    \begin{subfigure}[b]{\linewidth}
        \centering
        \includegraphics[width=\subfigwidth\linewidth,trim={2cm 2cm 2cm 3cm},clip,fbox]{diagrams/ros_scene_visualizations/case0/case0_cc_benign.png}
        \caption{Aggregator's perspective of unattacked scene.}
    \end{subfigure}
    \begin{subfigure}[b]{\linewidth}
        \centering
        \includegraphics[width=\subfigwidth\linewidth,trim={2cm 2cm 2cm 3cm},clip,fbox]{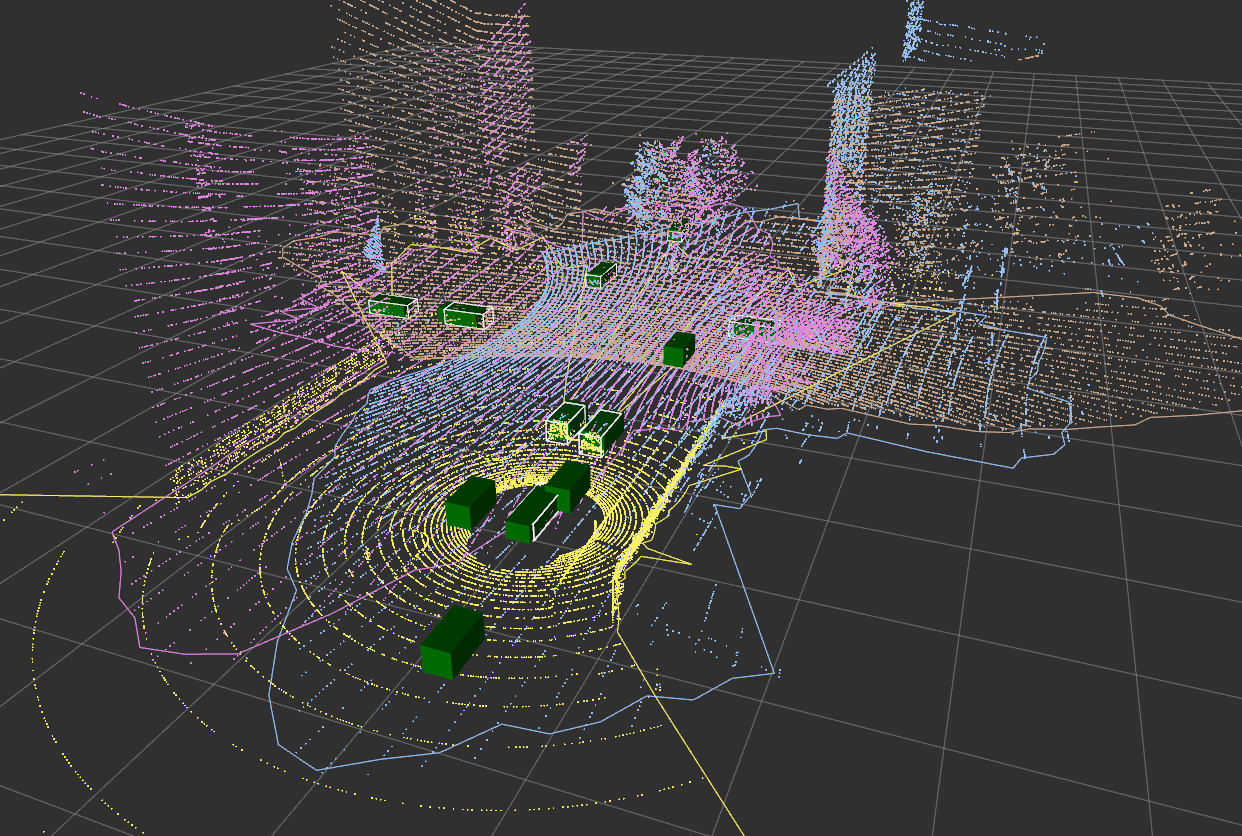}
        \caption{Compromised Agent 0 leads to malicious tracks at CC.}
    \end{subfigure}
    \caption{Each agent maintains local tracks and sends data to the aggregator (AGG). (a) In a benign case, AGG's green tracks match ground truth objects. (b) When Agent 0 is compromised by an FP attacker, without trust, spurious tracks are injected into AGG. With trust-aware fusion, the attack on Agent 0 is detected (Case 1 in Sec.~\ref{sec:exp-ue-dataset-case-results}). Data from agents in Cases 1-3 are shown in Figs.~\ref{fig:results-viz-case-0},~\ref{fig:results-viz-case-1}, and~\ref{fig:results-viz-case-2} in Appendix~\ref{appendix:results}.}
    \label{fig:multi-agent-ros-cc}
\end{figure}


\subsubsection{Baseline, benign dataset} \label{sec:exp-ue-dataset-baseline}

We build baseline scenes with mobile intelligent agents and static infrastructure sensors. Uncontrollable objects (NPCs) move in the scene via CARLA's autopilot. Intelligent agents with camera and \lidar\ sensors capture data at 10~Hz. The scenes are at \emph{safety-critical} intersections with occlusions; select frames are in Fig.~\ref{fig:case-diagrams}. We provide configuration files in our open-source release to aid in repeating dataset collection.

\subsubsection{Monte Carlo adversary datasets} \label{sec:exp-ue-dataset-adv}

The threat models from Sec.~\ref{sec:threat-model} are applied in postprocessing to the baseline datasets. Monte Carlo adversary randomization expands the set of benign scenes to hundreds of adversary conditions with uniquely compromised agents. We choose Monte Carlo randomization to not overfit the trust model to any hand-crafted case study. A sample of the parameters and their random functions is in Table~\ref{tab:dataset-mc}.

\begin{table}[t]
    \centering
    \begin{tabular}{c|c|c}
        \textbf{Parameter} & \textbf{Function} & \textbf{Inputs} \\
        \toprule \toprule
        Random seed & \texttt{RANDINT} & $[0,\, \infty]$ \\
        Baseline scene & \texttt{RANDCHOICE} & Available scenes \\
        Attacked agent & \texttt{RANDCHOICE} & $\{0,1,2,3\} $ \\
        \# FPs per adv. & \texttt{POISSON} & $\lambda=\lambda_{FP}$ \\
        \# FNs per adv. & \texttt{POISSON} & $\lambda=\lambda_{FN}$ \\
        Trans. dist, $d$ & \texttt{RANDN} & $\expectation[d]=\mu_d$, $\variance[d]=\sigma_d^2$ \\
        $\Delta t$ wait pre-attack & \texttt{RANDN} & $\expectation[\Delta t]=\mu_t$, $\variance[\Delta t]=\sigma_t^2$
        \end{tabular}
    \caption{A subset of the randomizable parameters for threat models from Sec.~\ref{sec:threat-model} available during Monte Carlo adversary dataset generation. The Monte Carlo randomization ensures the performance of security-aware sensor fusion is not tailored to one particular scene.}
    \label{tab:dataset-mc}
\end{table}

\subsubsection{Adversary case studies} \label{sec:exp-ue-dataset-cases}

We test security-aware fusion against adversarial situations and present detailed case study results. In Case 0, Agent 0 is subject to a static FP attacker. In Case 1, Agents 0 and 1 are both subject to Markovian FP attacks. Finally, in Case 2, Agent 3 is compromised by an FN attack. In Sec.~\ref{sec:exp-ue-dataset-case-results}, we explore the case study results in detail. Appendix~\ref{appendix:results} illustrates samples of data from the case studies in Figs.~\ref{fig:results-viz-case-0},~\ref{fig:results-viz-case-1}, and~\ref{fig:results-viz-case-2}. 

\subsection{Metrics} \label{sec:exp-ue-metrics}

Evaluating an MTT problem with and without attacks requires defining criteria to match tracks to truths and to compute data fusion performance given matches. We employ industry-standard methods for MTT evaluation with classical assignment-based metrics. We then propose novel trust metrics to capture performance of MATE within security-aware fusion. Derivation of novel metrics is in Appendix~\ref{appendix:metrics}.

\myparagraph{Assignment metrics.}{4pt}{0pt} Tracked objects either correspond to true objects or FPs. True objects are either matched with a track or are undetected (FNs). The classical metrics precision, recall, and F1-score (P, R, F1) are functions of the number of true objects, FPs, and FNs.

\myparagraph{OSPA metric.}{4pt}{0pt} Optimal Sub-Pattern Assignment (OSPA)~\cite{schuhmacher2008ospa} evaluates MTT. After matching tracks to truths, OSPA mixes estimation error and cardinality penalties (e.g., more truths than tracks) with smaller OSPA implying better performance.

\myparagraph{Novel track trust metric.}{4pt}{0pt} Track trust is a PDF on $[0,1]$. Optimal trust estimation has all density at 1 for a true object and at 0 for an FP. The estimator wants to maximize \emph{area above the trust cumulative distribution function} (CDF) for tracks on true objects and maximize \emph{area below the trust CDF} for tracks on FPs. The mean over all tracks summarizes the performance.

\myparagraph{Novel agent trust metric.}{4pt}{0pt} Similarly, agent trust is a probability density on $[0,1]$. Trust estimation should maximize area above the trust CDF for unattacked agents and should maximize area below when the agent is compromised.

\subsection{Monte Carlo Parameter Tuning} \label{sec:exp-ue-mctuning}

MATE and trust-informed fusion contain parameters including propagation time coefficients, trust thresholds, and agent/track negativity biases among others. To systematically evaluate parameter selections and to converge on a suitable configuration for testing, we run multiple trust models in parallel with Monte Carlo randomized parameters on top of the Monte Carlo adversarial datasets. The details of this tuning are in Appendix~\ref{appendix:results-mc-tuning}. The result of the tuning is a set of parameters to be used consistently for case study evaluation.

\subsection{Case Study Results} \label{sec:exp-ue-dataset-case-results}

We evaluate security-aware sensor fusion on case studies. In each case, we run both with and without security-aware fusion on both the benign (unattacked) and compromised scenes. On all combinations, we compute assignment and OSPA metrics. FPs will reduce precision while FNs will reduce recall; a translation attack will reduce both. Both FPs and FNs will increase OSPA. We then compute trust metrics on baseline and attacked scenes with security-aware fusion.

Videos of agent sensor, perception, and tracking data along with AGG fusion results are online at~\cite{mate-links}. We build a custom display to capture the evolution of trust data. Select snapshots and metrics timeseries are presented for each case in Figs. (\ref{fig:case-0-results},~\ref{fig:case-0-metrics}), (\ref{fig:case-1-results},~\ref{fig:case-1-metrics}), and (\ref{fig:case-2-results},~\ref{fig:case-2-metrics}) in the following sections. Supplemental visualizations are in Appendix~\ref{appendix:results-case-studies}.

\subsubsection{Case study 0: static FP attack on Agent 0}

\begin{figure}[t]
    \centering
    \begin{subfigure}[b]{\linewidth}
        \centering
        \includegraphics[width=\linewidth]{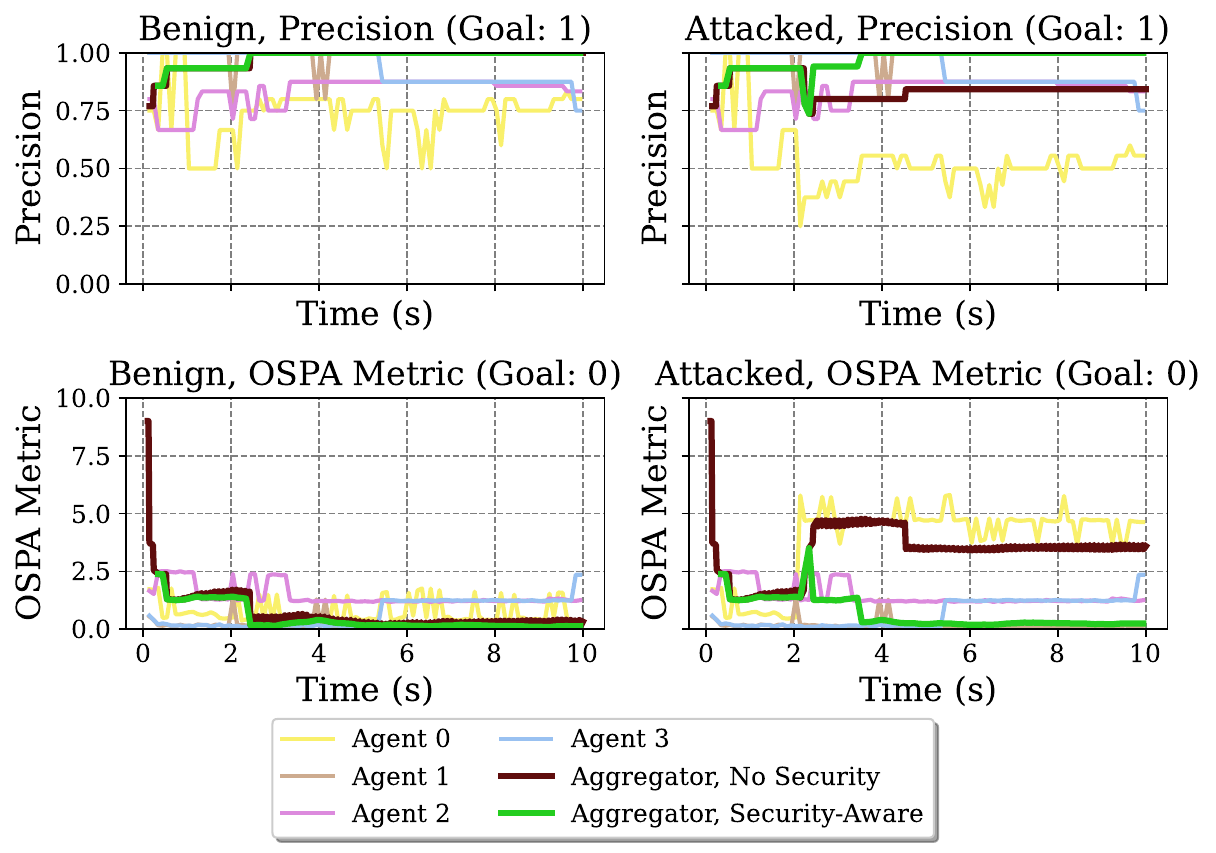}
        \caption{Sec.-aware returns prec./OSPA to baseline with adv. Agent0.}
        \label{fig:case-0-metrics-ospa}
    \end{subfigure}
    \begin{subfigure}[b]{\linewidth}
        \centering
        \includegraphics[width=\linewidth]{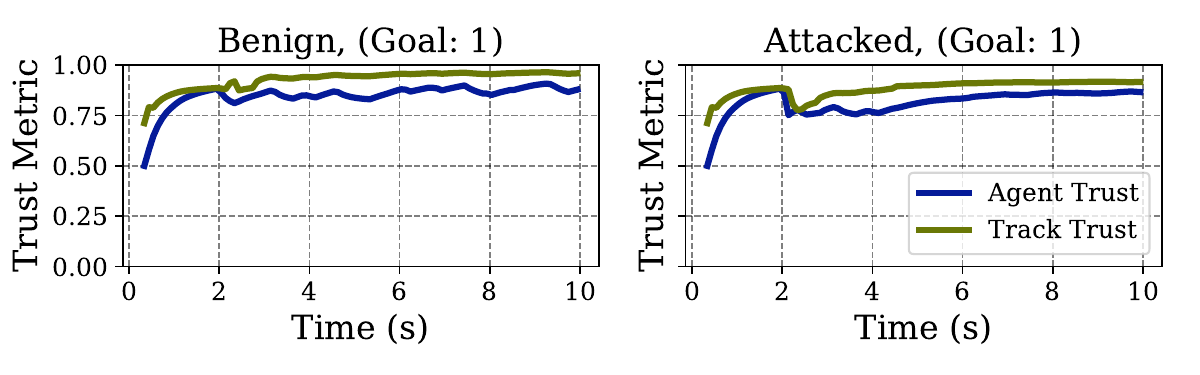}
        \caption{Trust estimation successful in benign and adversarial cases.}
        \label{fig:case-0-metrics-trust}
    \end{subfigure}
    \caption{(Case 0) (a) When Agent 0 is compromised via FP adversary, its OSPA metric increases relative to benign (yellow). This leads to an increase in the aggregator OSPA without security-awareness (red). Security awareness detects the FP attack and returns performance to the unattacked levels (green). (b) Over time, estimated trust distributions approach their ground truth targets in both benign and attacked cases. Agent 0 becomes adversarial at $t=2$ in ``Attacked'', leading to momentary discontinuity.}
    \label{fig:case-0-metrics}
\end{figure}

\begin{figure}[t]
    \centering
    \centering
    \includegraphics[width=\linewidth]{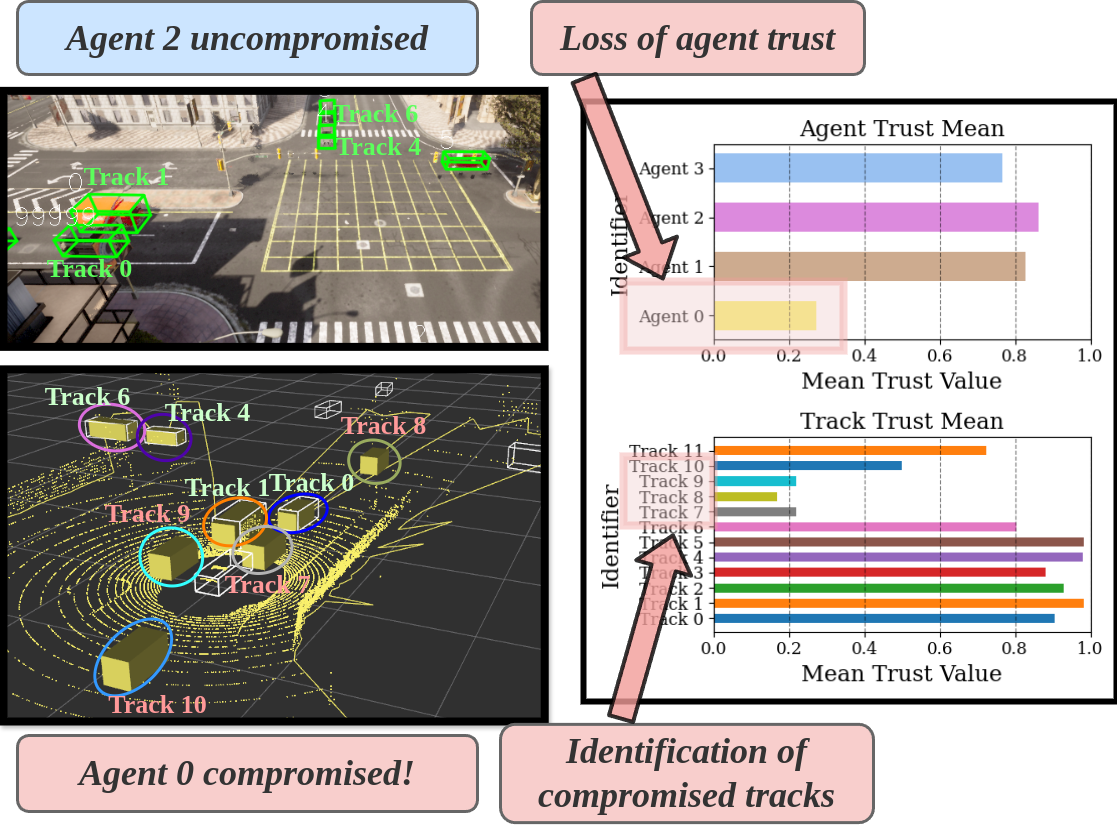}
    \caption{(Case 0) Agent 0 is compromised by a FP attacker. Adversary introduces malicious tracks into agent 0's set that agent 0 cannot detect alone. Through collaborative security-aware fusion, new tracks are identified as adversarial and agent 0's trust is appropriately degraded.}
    \label{fig:case-0-results}
\end{figure}

Agent 0's detections are compromised by a static FP attacker. In Agent 0's local tracking, the FPs propagate into adversarial tracks, and OSPA increases compared to the benign case, as in Fig.~\ref{fig:case-0-metrics-ospa}. Without security, FP tracks are accepted by AGG and OSPA is compromised, as in Fig.~\ref{fig:case-0-metrics-ospa}. With MATE driving trust-informed fusion, false tracks in Agent 0 are detected and given low trust score, as in Fig.~\ref{fig:case-0-results}. Similarly, the trust on Agent 0 decreases. Informing fusion of the distrustedness of agents/objects allows for filtering FPs and a return of the trust-informed OSPA to baseline levels. Security-aware fusion leads to a 94\% reduction in OSPA error from the attacked, no-security case relative to baseline. Moreover, in Fig.~\ref{fig:case-0-metrics-trust}, mean agent/track trust metrics of 87\% and 92\% indicate that trust estimation adequately identifies false/distrusted tracks/agents.


\subsubsection{Case study 1: Markov FP attacks on Agents 0, 1}
\begin{figure}[t]
    \centering
    \centering
    \includegraphics[width=0.8\linewidth]{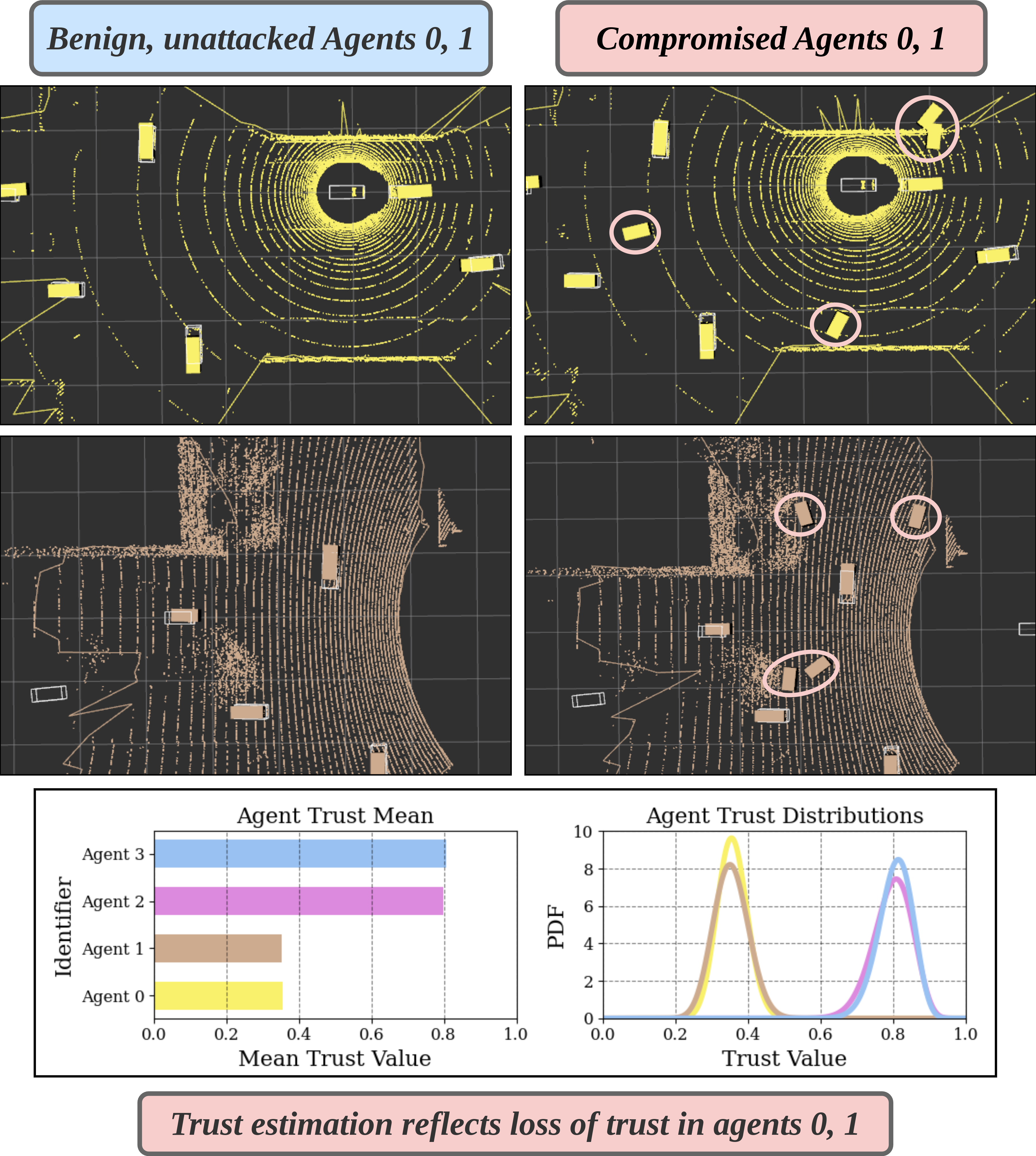}
    \caption{In case 1, half of agents are compromised (agents 0, 1). Nevertheless, trust estimation identifies agents 0, 1 are providing malicious information rapidly and accurately. (Left) shows baseline perception and tracking performance on benign data while (right) shows compromised outcomes.}
    \label{fig:case-1-results}
\end{figure}

\begin{figure}[t]
    \centering
    \begin{subfigure}[b]{\linewidth}
        \centering
        \includegraphics[width=\linewidth]{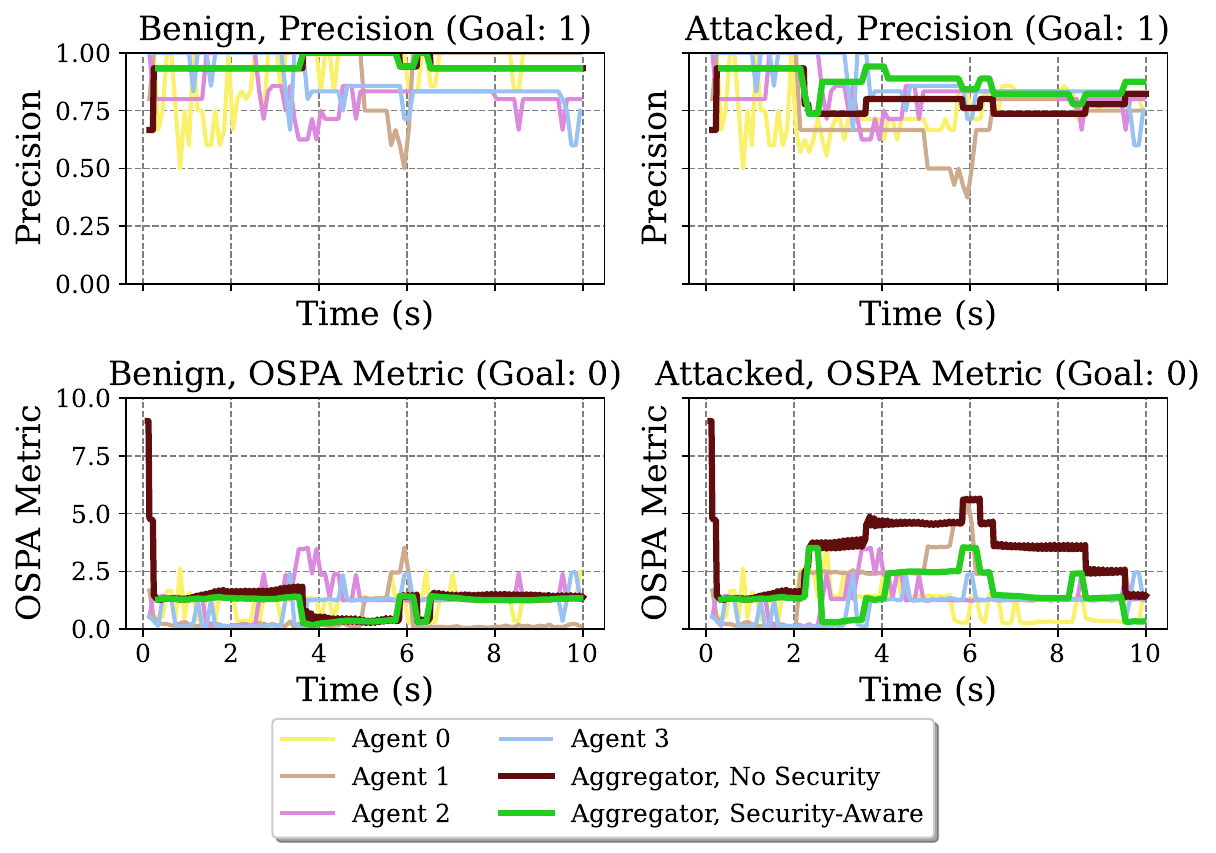}
        \caption{Attack on Agents 0, 1 leads to dramatic increase in OSPA without Sec. Sec-aware improves outcomes at the aggregator.}
        \label{fig:case-1-metrics-ospa}
    \end{subfigure}
    \begin{subfigure}[b]{\linewidth}
        \centering
        \includegraphics[width=\linewidth]{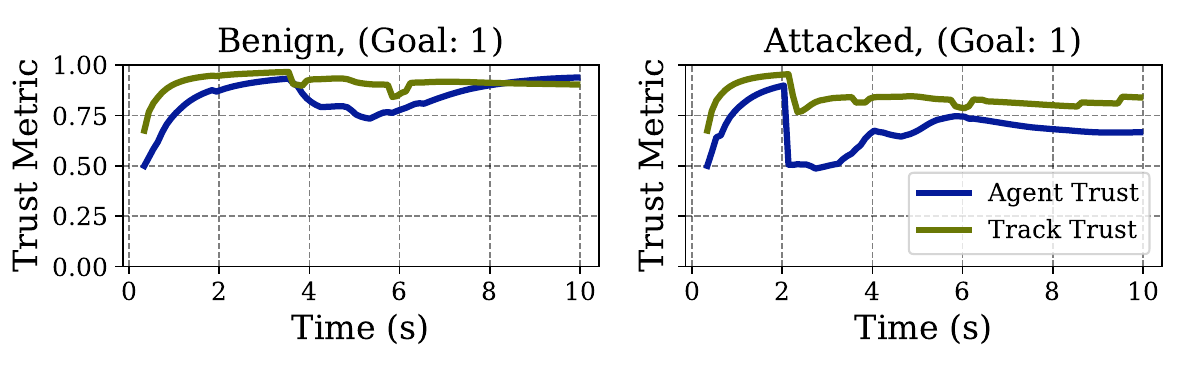}
        \caption{Trust estimation successful in benign and adversarial cases.}
        \label{fig:case-1-metrics-trust}
    \end{subfigure}
    \caption{(Case 1) (a) Agents 0 and 1 are attacked by FP adversaries started at $t=2s$ leading to dramatic increases in their OSPA error. Attacks affect a non-security-aware fusion algorithm and lead to undesirably high levels of OSPA error (red). With security-awareness (green), FPs are identified and removed, leading to a reduction in OSPA error and levels on-par with same scene without attacks (baseline). (b) Estimated trust states are accurate relative to ground truth. Agent target (trusted/distrusted) flips once attack commences at $t=2s$.}
    \label{fig:case-1-metrics}
\end{figure}

Both Agent 0's and Agent 1's detections are compromised by Markovian FP attackers. Despite the dynamic propagation of tracks, Fig.~\ref{fig:case-1-results} highlights that security-aware sensor fusion detects malicious tracks and identifies Agents 0 and 1 as distrusted. A significant number of adversarial false tracks deteriorates non-secure sensor fusion. OSPA increases for Agents 0 and 1 from near an average score of 1.2 in the benign case to over 7 in the attacked circumstance, as in Fig.~\ref{fig:case-1-metrics-ospa}. Eliminating the false tracks with security-awareness returns the OSPA to near-baseline levels with an average reduction of 76\%. Similarly, track and agent trust are consistently high, indicating accurate and responsive trust estimation.


\begin{figure}[t]
    \centering
    \centering
    \includegraphics[width=\linewidth]{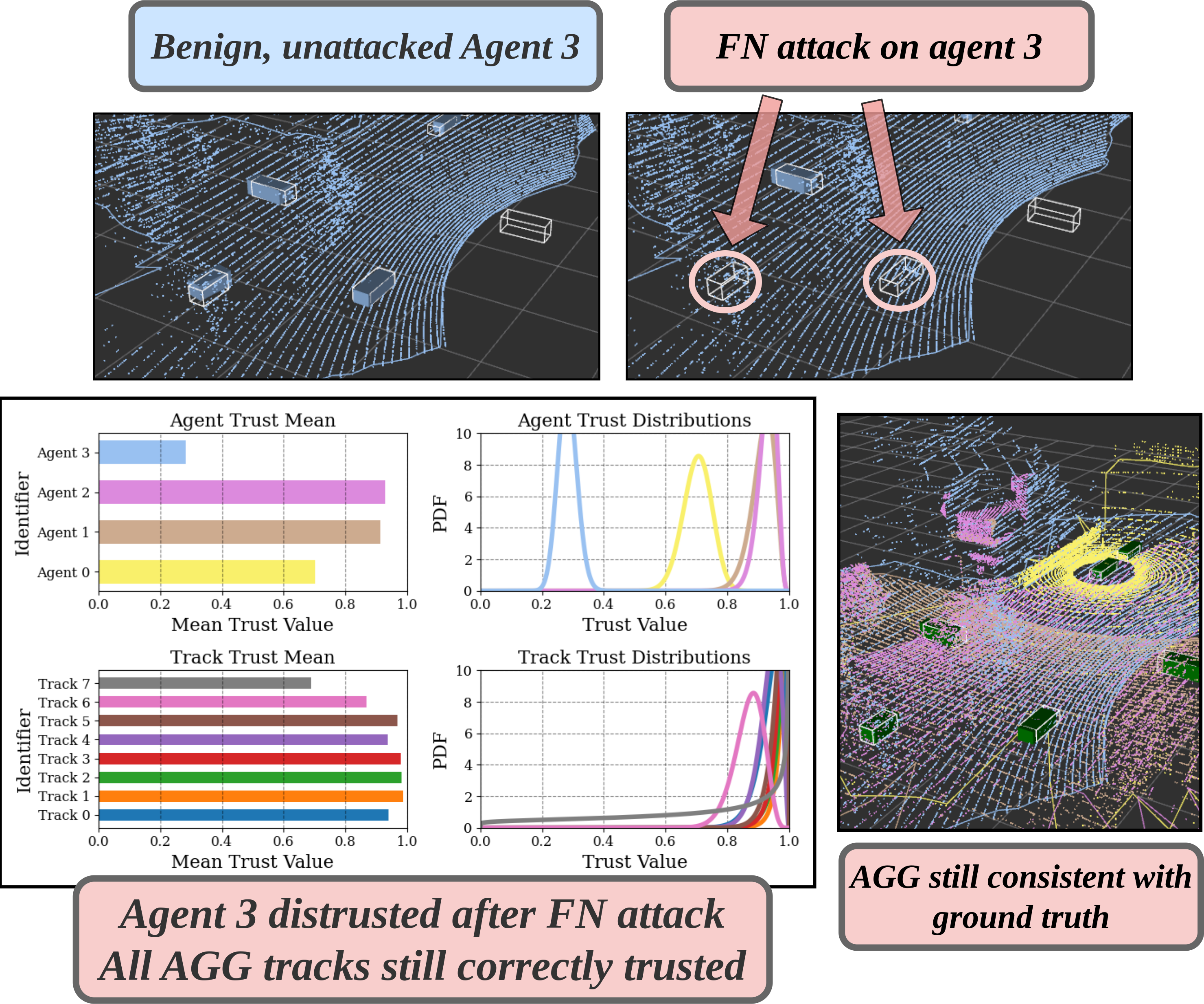}
    \caption{Case 2 sees Agent 3 compromised by a FN attack on just two of the agent's tracked objects. Agent 3 is quickly registered as distrusted. All tracks at AGG still have high trust because they are not compromised.}
    \label{fig:case-2-results}
\end{figure}
\begin{figure}[!t]
    \centering
    \begin{subfigure}[b]{\linewidth}
        \centering
        \includegraphics[width=\linewidth]{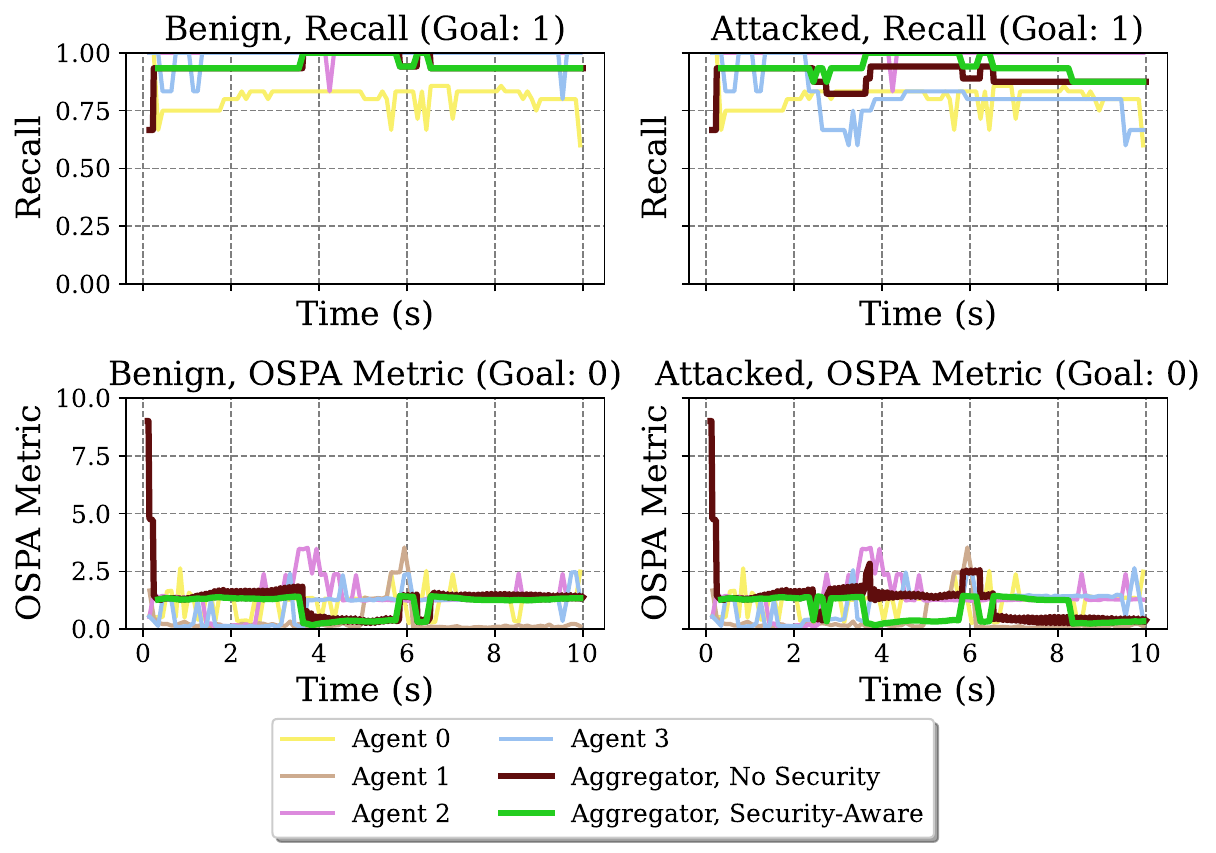}
        \caption{Attack on Agents 0, 1 leads to dramatic increase in OSPA without Sec. Sec-aware improves outcomes at the aggregator.}
        \label{fig:case-2-metrics-ospa}
    \end{subfigure}
    \begin{subfigure}[b]{\linewidth}
        \centering
        \includegraphics[width=\linewidth]{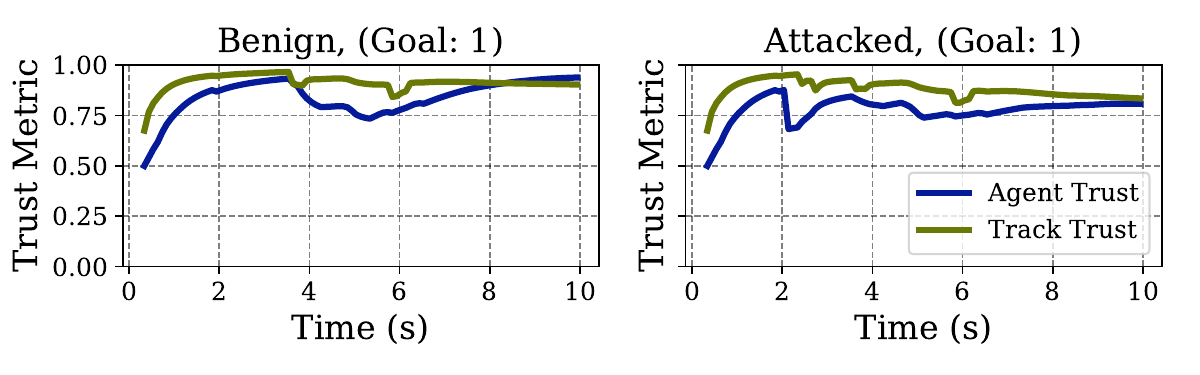}
        \caption{Trust estimation successful in benign and adversarial cases.}
        \label{fig:case-2-metrics-trust}
    \end{subfigure}
    \caption{(Case 2) (a) Agent 3 is compromised by a small number of FN attacks at $t=2s$. Average OSPA metric does not significantly change due to the attack affecting only a small number of objects. FN attack on a single agent does not alter the OSPA metric at the AGG because other agents can compensate for the missing detections. (b) Agent and track trust are still estimated correctly despite the attack being small in magnitude.}
    \label{fig:case-2-metrics}
\end{figure}

\subsubsection{Case study 2: FN attack on Agent 3}
Agent 3's detections are compromised by an FN attacker. Two tracks are consistently invalidated at Agent 3's local perception leading to an increase in OSPA, as in Fig.~\ref{fig:case-2-metrics-ospa}. FN attacks on small numbers of agents may not impact OSPA at AGG, even without security-aware fusion, because other agents compensate if they have overlapping fields of view. However, the compromised Agent 3's recall decreases, as in Fig.~\ref{fig:case-2-metrics-ospa}. With security awareness, despite being a small perturbation that does not affect AGG performance metrics, the Agent 3 is quickly identified as distrusted, as in Fig.~\ref{fig:case-2-results}. Trust is estimated consistently throughout the experiment with agent and track trust metrics at high levels, as in Fig.~\ref{fig:case-2-metrics-trust}. 

\subsection{Near-Real-Time Autonomy} \label{sec:exp-ue-ros}

Running perception, tracking, and security-aware fusion in a single process is limited by global interpreter lock (GIL) that prohibits multiple threads/processes from fully leveraging compute resources. To facilitate efficient algorithms that scale with increased number of agents, we implement a full-stack multi-agent pipeline (i.e.,~perception, tracking, fusion, security) in the robot operating system (ROS)~\cite{quigley2009ros}. Data from cameras and LiDARs stream over topics to perception nodes and inference results are sent to trackers that filter object states. Agent-local results are sent to AGG. All of these computationally intensive algorithms run as ROS nodes to best utilize available resources in near-real time.

For efficient analysis in postprocessing, case study datasets are converted to ROS bags that can be replayed at fixed rates - either slower or faster than ``real-time'' to scale to the compute of the host machine. This is made possible by \avstack's integrations with OpenMMLab~\cite{chen2019mmdetection} and ROS.

\section{Discussion} \label{sec:discussion}

With sufficient resources, an attacker can thwart many algorithms; trust-based sensor fusion is no exception. We discuss several ways for attackers to compromise security-aware fusion. These components ought to be given significant attention before deploying in contested environments. 

\myparagraph{Attack coordination.}{4pt}{0pt} If multiple attackers can consistently coordinate, then e.g., a FP could be reinforced by multiple agents simultaneously, increasing the likelihood of inaccurate PSMs. It is more challenging to identify a FP that is consistently detected by many agents unless a subset of negatively reporting agents are strongly trusted. 

\myparagraph{Compromised FOV.}{4pt}{0pt} If an adversary can compromise detection messages, he can compromise FOV messages. Analysis of compromised FOV was out of scope of this work, but it will result in disagreements between agents and impact agent/track trust. This suggests trust estimation can detect FOV attacks. Future works will analyze compromised FOV and will consider certifiable robustness of FOV estimation.

\myparagraph{Low density of agents.}{4pt}{0pt} Low density coverage leaves the trust model ineffective; the veracity of a track cannot be determined from a single agent. Similarly, if the overlapping space between two agent FOVs yields no tracked objects, the model cannot derive useful PSMs to update agent trust. Limited trust observability is likely to occur if agent density is low.
\section{Conclusion and Future Work} \label{sec:conclusion}

We proposed novel algorithms for security-aware sensor fusion in multi-agent autonomy. Our method maps perception-oriented sensor data and agent pose to the space of \emph{trust} and estimates PDFs of the trustedness of agents and their tracked objects. Trust estimates feed trust-informed data fusion that filters distrusted information to provide secure and assured situational awareness. Our approach to security-aware sensor fusion mitigates state of the art FP, FN, and translation attacks in safety-critical scenarios in urban environments while maintaining strong performance in an unattacked baseline. 

We plan model improvements in the following areas. (1) An attractive property of our Bayesian approach is the ability to specify priors on agent and track trustedness. Due to space limitations, we could not fully explore prior information in this work; informative priors will be investigated in future works. (2) We formed PSM functions using assignment-based logic. Considering alternative definitions that leverage more complex inputs such as probabilistic observations and consistency of negative space (i.e.,~absence of detections) may improve trust estimation performance. (3) Finally, trust posterior estimation was made tractable by independence assumption and Gibbs sampling. While experimental analysis suggest minimal performance loss, future works will investigate nonparametric numerical methods of posterior estimation.


\section*{Statements}

\myparagraph{Ethics considerations.}{4pt}{0pt} This work proposes defenses against attacks on AVs and raises minimal ethical considerations. All research was conducted using datasets and simulations.

\myparagraph{Open science.}{4pt}{0pt} As part of the compliance to the open science initiative, we will release all datasets, models, and source code used to derive the results in this work. The source code will be released online for researchers to reproduce our outcomes.

\bibliographystyle{IEEEtranS}
\bibliography{references}

\begin{thebibliography}{10}
\providecommand{\url}[1]{#1}
\csname url@samestyle\endcsname
\providecommand{\newblock}{\relax}
\providecommand{\bibinfo}[2]{#2}
\providecommand{\BIBentrySTDinterwordspacing}{\spaceskip=0pt\relax}
\providecommand{\BIBentryALTinterwordstretchfactor}{4}
\providecommand{\BIBentryALTinterwordspacing}{\spaceskip=\fontdimen2\font plus
\BIBentryALTinterwordstretchfactor\fontdimen3\font minus \fontdimen4\font\relax}
\providecommand{\BIBforeignlanguage}[2]{{%
\expandafter\ifx\csname l@#1\endcsname\relax
\typeout{** WARNING: IEEEtranS.bst: No hyphenation pattern has been}%
\typeout{** loaded for the language `#1'. Using the pattern for}%
\typeout{** the default language instead.}%
\else
\language=\csname l@#1\endcsname
\fi
#2}}
\providecommand{\BIBdecl}{\relax}
\BIBdecl

\bibitem{allig2019trustworthiness}
C.~Allig, T.~Leinm{\"u}ller, P.~Mittal, and G.~Wanielik, ``Trustworthiness estimation of entities within collective perception,'' in \emph{2019 IEEE Vehicular Networking Conference (VNC)}.\hskip 1em plus 0.5em minus 0.4em\relax IEEE, 2019, pp. 1--8.

\bibitem{ambrosin2019design}
M.~Ambrosin, L.~L. Yang, X.~Liu, M.~R. Sastry, and I.~J. Alvarez, ``Design of a misbehavior detection system for objects based shared perception v2x applications,'' in \emph{2019 IEEE Intelligent Transportation Systems Conference (ITSC)}.\hskip 1em plus 0.5em minus 0.4em\relax IEEE, 2019, pp. 1165--1172.

\bibitem{mate-links}
anonymous authors, ``{Security-Aware Sensor Fusion},'' \url{https://sites.google.com/view/trusted-sensor-fusion/}, 2024.

\bibitem{ansari2021v2x}
M.~R. Ansari, J.-P. Monteuuis, J.~Petit, and C.~Chen, ``V2x misbehavior and collective perception service: Considerations for standardization,'' in \emph{2021 IEEE Conference on Standards for Communications and Networking (CSCN)}.\hskip 1em plus 0.5em minus 0.4em\relax IEEE, 2021, pp. 1--6.

\bibitem{athalye2018obfuscated}
A.~Athalye, N.~Carlini, and D.~Wagner, ``Obfuscated gradients give a false sense of security: Circumventing defenses to adversarial examples,'' in \emph{International conference on machine learning}.\hskip 1em plus 0.5em minus 0.4em\relax PMLR, 2018, pp. 274--283.

\bibitem{bar2004estimation}
Y.~Bar-Shalom, X.~R. Li, and T.~Kirubarajan, \emph{Estimation with applications to tracking and navigation: theory algorithms and software}.\hskip 1em plus 0.5em minus 0.4em\relax John Wiley \& Sons, 2004.

\bibitem{bar1995multitarget}
Y.~Bar-Shalom and X.-R. Li, \emph{Multitarget-multisensor tracking: principles and techniques}.\hskip 1em plus 0.5em minus 0.4em\relax YBS publishing Storrs, CT, 1995, vol.~19.

\bibitem{bissmeyer2012assessment}
N.~Bi{\ss}meyer, S.~Mauthofer, K.~M. Bayarou, and F.~Kargl, ``Assessment of node trustworthiness in vanets using data plausibility checks with particle filters,'' in \emph{2012 IEEE Vehicular Networking Conference (VNC)}.\hskip 1em plus 0.5em minus 0.4em\relax IEEE, 2012, pp. 78--85.

\bibitem{1986blackmanRadar}
S.~S. Blackman, ``Multiple-target tracking with radar applications,'' \emph{Dedham}, 1986.

\bibitem{2019cao-spoofing}
Y.~Cao, C.~Xiao, B.~Cyr, Y.~Zhou, W.~Park, S.~Rampazzi, Q.~A. Chen, K.~Fu, and Z.~M. Mao, ``Adversarial sensor attack on lidar-based perception in autonomous driving,'' in \emph{2019 ACM CCS}.\hskip 1em plus 0.5em minus 0.4em\relax London, UK: ACM, 2019, pp. 2267--2281.

\bibitem{casella1992explaining}
G.~Casella and E.~I. George, ``Explaining the gibbs sampler,'' \emph{The American Statistician}, vol.~46, no.~3, pp. 167--174, 1992.

\bibitem{chen2019mmdetection}
K.~Chen, J.~Wang, J.~Pang, Y.~Cao, Y.~Xiong, X.~Li, S.~Sun, W.~Feng, Z.~Liu, J.~Xu \emph{et~al.}, ``Mmdetection: Open mmlab detection toolbox and benchmark,'' \emph{arXiv preprint arXiv:1906.07155}, 2019.

\bibitem{cook1984distributed}
R.~L. Cook, T.~Porter, and L.~Carpenter, ``Distributed ray tracing,'' in \emph{Proceedings of the 11th annual conference on Computer graphics and interactive techniques}, 1984, pp. 137--145.

\bibitem{crouse2016implementing}
D.~F. Crouse, ``On implementing 2d rectangular assignment algorithms,'' \emph{IEEE Transactions on Aerospace and Electronic Systems}, vol.~52, no.~4, pp. 1679--1696, 2016.

\bibitem{2017carla}
A.~Dosovitskiy, G.~Ros, F.~Codevilla, A.~Lopez, and V.~Koltun, ``Carla: An open urban driving simulator,'' in \emph{CoRL}.\hskip 1em plus 0.5em minus 0.4em\relax PMLR, 2017, pp. 1--16.

\bibitem{eykholt2018robust}
K.~Eykholt, I.~Evtimov, E.~Fernandes, B.~Li, A.~Rahmati, C.~Xiao, A.~Prakash, T.~Kohno, and D.~Song, ``Robust physical-world attacks on deep learning visual classification,'' in \emph{Proceedings of the IEEE CVPR}, 2018, pp. 1625--1634.

\bibitem{golle2004detecting}
P.~Golle, D.~Greene, and J.~Staddon, ``Detecting and correcting malicious data in vanets,'' in \emph{Proceedings of the 1st ACM international workshop on Vehicular ad hoc networks}, 2004, pp. 29--37.

\bibitem{2022hally-frustum}
R.~S. Hallyburton, Y.~Liu, Y.~Cao, Z.~M. Mao, and M.~Pajic, ``Security analysis of camera-lidar fusion against black-box attacks on autonomous vehicles,'' in \emph{31st USENIX SECURITY)}.\hskip 1em plus 0.5em minus 0.4em\relax Berkeley, CA: USENIX, 2022, pp. 1--18.

\bibitem{hallyburton2023datasets}
R.~S. Hallyburton and M.~Pajic, ``Datasets, models, and algorithms for multi-sensor, multi-agent autonomy using avstack,'' \emph{arXiv preprint arXiv:2312.04970}, 2023.

\bibitem{hallyburton2024bayesian}
------, ``Bayesian methods for trust in collaborative multi-agent autonomy,'' \emph{arXiv preprint arXiv:2403.16956}, 2024.

\bibitem{hallyburton2023partial}
R.~S. Hallyburton, Q.~Zhang, Z.~M. Mao, and M.~Pajic, ``Partial-information, longitudinal cyber attacks on lidar in autonomous vehicles,'' \emph{arXiv preprint arXiv:2303.03470}, 2023.

\bibitem{hallyburton2023avstack}
R.~S. Hallyburton, S.~Zhang, and M.~Pajic, ``Avstack: An open-source, reconfigurable platform for autonomous vehicle development,'' in \emph{ACM/IEEE ICCPS}, 2023, pp. 209--220.

\bibitem{huhns2002trusted}
M.~N. Huhns and D.~A. Buell, ``Trusted autonomy,'' \emph{IEEE Internet Computing}, vol.~6, no.~3, p.~92, 2002.

\bibitem{jia2020fooling}
Y.~J. Jia, Y.~Lu, J.~Shen, Q.~A. Chen, H.~Chen, Z.~Zhong, and T.~W. Wei, ``Fooling detection alone is not enough: Adversarial attack against multiple object tracking,'' in \emph{International Conference on Learning Representations (ICLR'20)}, 2020.

\bibitem{kwon2014proactive}
J.~Kwon and M.~E. Johnson, ``Proactive versus reactive security investments in the healthcare sector,'' \emph{Mis Quarterly}, vol.~38, no.~2, pp. 451--A3, 2014.

\bibitem{lang2019pointpillars}
A.~H. Lang, S.~Vora, H.~Caesar, L.~Zhou, J.~Yang, and O.~Beijbom, ``Pointpillars: Fast encoders for object detection from point clouds,'' in \emph{Proceedings of the IEEE/CVF CVPR}, 2019, pp. 12\,697--12\,705.

\bibitem{liu2021seeing}
J.~Liu and J.-M. Park, ``“seeing is not always believing”: detecting perception error attacks against autonomous vehicles,'' \emph{IEEE Transactions on Dependable and Secure Computing}, vol.~18, no.~5, pp. 2209--2223, 2021.

\bibitem{liu2021miso}
X.~Liu, L.~Yang, ..., and L.~G. Baltar, ``Miso-v: Misbehavior detection for collective perception services in vehicular communications,'' in \emph{2021 IEEE IV}.\hskip 1em plus 0.5em minus 0.4em\relax IEEE, 2021, pp. 369--376.

\bibitem{lo2007illusion}
N.-W. Lo and H.-C. Tsai, ``Illusion attack on vanet applications-a message plausibility problem,'' in \emph{2007 IEEE globecom workshops}.\hskip 1em plus 0.5em minus 0.4em\relax IEEE, 2007, pp. 1--8.

\bibitem{monteuuis2018my}
J.~Monteuuis, J.~Petit, ..., S.~Mafrica, and A.~Servel, ``“my autonomous car is an elephant”: A machine learning based detector for implausible dimension,'' in \emph{2018 IEEE SSIC}, 2018, pp. 1--8.

\bibitem{pajic2014robustness}
M.~Pajic, J.~Weimer, N.~Bezzo, P.~Tabuada, O.~Sokolsky, I.~Lee, and G.~J. Pappas, ``Robustness of attack-resilient state estimators,'' in \emph{ACM/IEEE ICCPS}, 2014, pp. 163--174.

\bibitem{park2012newconcave}
J.-S. Park and S.-J. Oh, ``A new concave hull algorithm and concaveness measure for n-dimensional datasets,'' \emph{Journal of Information science and engineering}, vol.~28, no.~3, pp. 587--600, 2012.

\bibitem{2014cyberattacks}
J.~Petit and S.~E. Shladover, ``Potential cyberattacks on automated vehicles,'' \emph{IEEE Transactions on Intelligent transportation systems}, vol.~16, no.~2, pp. 546--556, 2014.

\bibitem{quigley2009ros}
M.~Quigley, K.~Conley, B.~Gerkey, J.~Faust, T.~Foote, J.~Leibs, R.~Wheeler, A.~Y. Ng \emph{et~al.}, ``Ros: an open-source robot operating system,'' in \emph{ICRA workshop}, vol.~3, no. 3.2.\hskip 1em plus 0.5em minus 0.4em\relax Kobe, Japan, 2009, p.~5.

\bibitem{schuhmacher2008ospa}
D.~Schuhmacher, B.-T. Vo, and B.-N. Vo, ``A consistent metric for performance evaluation of multi-object filters,'' \emph{IEEE transactions on signal processing}, vol.~56, no.~8, pp. 3447--3457, 2008.

\bibitem{shah2011rumors}
D.~Shah and T.~Zaman, ``Rumors in a network: Who's the culprit?'' \emph{IEEE Transactions on information theory}, vol.~57, no.~8, pp. 5163--5181, 2011.

\bibitem{soleymani2017secure}
S.~A. Soleymani, A.~H. Abdullah, M.~Zareei, M.~H. Anisi, C.~Vargas-Rosales, M.~K. Khan, and S.~Goudarzi, ``A secure trust model based on fuzzy logic in vehicular ad hoc networks with fog computing,'' \emph{IEEE Access}, vol.~5, pp. 15\,619--15\,629, 2017.

\bibitem{sun2020towards}
J.~Sun, Y.~Cao, Q.~A. Chen, and Z.~M. Mao, ``Towards robust $\{$LiDAR-based$\}$ perception in autonomous driving: General black-box adversarial sensor attack and countermeasures,'' in \emph{29th USENIX Security Symposium (USENIX Security 20)}, 2020, pp. 877--894.

\bibitem{theodorakopoulos2004trust}
G.~Theodorakopoulos and J.~S. Baras, ``Trust evaluation in ad-hoc networks,'' in \emph{Proceedings of the 3rd ACM workshop on Wireless security}, 2004, pp. 1--10.

\bibitem{thys2019fooling}
S.~Thys, W.~Van~Ranst, and T.~Goedem{\'e}, ``Fooling automated surveillance cameras: adversarial patches to attack person detection,'' in \emph{Proceedings of the IEEE/CVF CVPR workshops}, 2019, pp. 0--0.

\bibitem{tramer2020adaptive}
F.~Tramer, N.~Carlini, W.~Brendel, and A.~Madry, ``On adaptive attacks to adversarial example defenses,'' \emph{Advances in neural information processing systems}, vol.~33, pp. 1633--1645, 2020.

\bibitem{tsukada2022misbehavior}
M.~Tsukada, S.~Arii, H.~Ochiai, and H.~Esaki, ``Misbehavior detection using collective perception under privacy considerations,'' in \emph{2022 IEEE CCNC}.\hskip 1em plus 0.5em minus 0.4em\relax IEEE, 2022, pp. 808--814.

\bibitem{van2018survey}
R.~W. Van Der~Heijden, S.~Dietzel, T.~Leinm{\"u}ller, and F.~Kargl, ``Survey on misbehavior detection in cooperative intelligent transportation systems,'' \emph{IEEE Communications Surveys \& Tutorials}, vol.~21, no.~1, pp. 779--811, 2018.

\bibitem{wang2006autonomous}
W.~Wang, G.~Zeng, and T.~Liu, ``An autonomous trust construction system based on bayesian method,'' in \emph{2006 IEEE/WIC/ACM International Conference on Intelligent Agent Technology}.\hskip 1em plus 0.5em minus 0.4em\relax IEEE, 2006, pp. 357--362.

\bibitem{yeremenko2017secure}
O.~Yeremenko, O.~Lemeshko, and A.~Persikov, ``Secure routing in reliable networks: proactive and reactive approach,'' in \emph{Conference on Computer Science and Information Technologies}.\hskip 1em plus 0.5em minus 0.4em\relax Springer, 2017, pp. 631--655.

\bibitem{yue2007intrusion}
W.~T. Yue and M.~Cakanyildirim, ``Intrusion prevention in information systems: Reactive and proactive responses,'' \emph{Journal of Management Information Systems}, vol.~24, no.~1, pp. 329--353, 2007.

\bibitem{zhang2024data}
Q.~Zhang, ..., Q.~A. Chen, and Z.~M. Mao, ``On data fabrication in collaborative vehicular perception: Attacks and countermeasures,'' in \emph{33rd USENIX Security}, 2024, pp. 6309--6326.

\end{thebibliography}

\appendix
\section{Collaborative Sensor Fusion}\label{appendix:fusion}

\subsection{Multi-Agent MTT}
We assume that agents $k=1,...,K$ provide $Q_{k,t} \geq 0$ detections at time $t$. We use $Z_t\coloneqq\{z_{q,k,t}\}|_{q\in[1,...,Q_{k,t}],k=1,...,K}$ as the set of the \emph{set of detections} from all agents and $X_t\coloneqq \{x_{i,t}\}$ as the set of all $N_t$ true object states. Formally, the objective of MTT is to estimate the joint posterior:
\begin{align}
    \Pr(X_t | Z_{1:t}) = \frac{\Pr(Z_t|X_t) \Pr(X_t|Z_{1:t-1})}{\Pr(Z_t | Z_{1:t-1})},
\end{align}
where $\Pr$ is a probability density. At each step, MTT retains a set of tracks, $\hat{X}_t \coloneqq \{\hat{x}_{j,t}\}$ as estimates of object states. Subscripts $j$ do not necessarily align with $i$ since $\hat{X_t}$ estimates both existence and state, e.g.,~$\hat{X}_t$ can have natural FPs or FNs and both $\hat{X}_t,\, X_t$ are permutation-invariant. 

MTT usually takes a two-stage approach to reduce the multi-object posterior to multiple single-object problems. Instead of using all measurements to update all tracks, MTT often assigns measurements to specific tracks for single-track updating (see many examples in~\cite{1986blackmanRadar,bar1995multitarget}). Steps include:
\begin{enumerate}
    \item \textbf{Data association:} perform bipartite matching to assign current detections, $Z_t$, to estimated track states, $\hat{X}_{t-1}$. Often, a measurement can only be used for a single track. Detections without a track start new tracks, tracks without detections are considered ``missed''.
    \item \textbf{Existence \& state estimation:} for each track, use assigned measurements from data association to update the track existence probability and state estimate.
\end{enumerate}

The measurements help the existence task reason whether tracks represent real objects or are FPs. The state estimation task employs an estimator such as the Kalman filter to mix measurements and kinematic models. Important to MTT is both agent pose (i.e.,~position and orientation) and the field of view (FOV) model that takes as input a point in space and determines if agent $k$ could reasonably observe an object at that point, if there existed one. The FOV model is important e.g.,~so as not to penalize agents and tracks for ``misses' when the candidate track was not expected to be visible to the agent. We group both under the term ``agent characteristics'', $A_t\coloneqq \{a_{k,t}\}$, and assume $A_t$ is known and uncompromised.

Local filtering is important for accounting for unavoidable errors due to natural uncertainties from real-world perception data. Tracking is known to be resilient to FPs and FNs arising from short-term errors in sensors~\cite{1986blackmanRadar}. Filtering may be able to stop some adversarial attacks, however, it is well-known that worst-case attackers manipulating objects consistent with plausible dynamics models will bypass such filtering methods~\cite{2022hally-frustum}. This fact motivates our implementation of security-aware sensor fusion to detect such challenging attacks leveraging multi-agent collaboration. 

\subsection{Field of View Modeling} \label{appendix:mate-fov}

Generating PSMs in the trust model relies on an accurate FOV to determine the set of objects an agent was \emph{expected} to see. Unfortunately, few FOV estimation algorithms exist in the literature.Ray tracing outperforms hull-based estimation in all tests. For ray tracing, LiDAR data is projected onto the ground plane, centered at the point cloud centroid, and mapped into polar coordinates (range, azimuth) in BEV. This polar bijection between azimuth and range prevents crossed FOV shapes, as points cannot pass through walls. We test both quantized and continuous ray tracing. To determine the FOV, we build FOV functions applied to LiDAR point clouds in BEV (top-down) and output a bounding polygon that describes the visible space.

\myparagraph{Ray tracing.}{4pt}{0pt} LiDAR data is projected into the ground plane, centered about the LiDAR point cloud's centroid, and projected into polar coordinates in the bird's eye view (BEV) (range, azimuth). Ray tracing leverages the bijection between azimuth angle and range for LiDAR point clouds in polar coordinates; e.g., points cannot travel through walls to obtain crossed-over FOV shapes. The bijection, upon polar projection, allows ray tracing to be highly effective at FOV estimation. Each angle maps to a range that represents the maximal point returned at that angle. We test quantized and continuous versions of ray tracing.

\myparagraph{Concave hull estimation.}{4pt}{0pt} Concave hull estimation is suitable FOV fitting without the bijection assumption. We use an algorithm off-the-shelf from~\cite{park2012newconcave}. The concave hull algorithm fits a polygon to the \lidar\ point cloud to approximate the visible space in the BEV reference frame.

\section{Supplemental MATE Derivation} \label{appendix:mate}

\subsection{Iterative updating with conditionals} \label{appendix:mate-gibbs}

\subsubsection{Gibbs Sampling}
During measurement generation, we obtain PSMs for agents and tracks. The full trust distribution is a complex multivariate with cross-correlations. This makes it difficult to sample from/estimate the parameters of the joint trust distribution. Luckily, the conditionals
\begin{equation}
\begin{aligned}
    (1)\ &\text{Track trust:} \ \Pr(\Tau^c_{t}\ |\ \Tau^a_{t-1}, Z_{1:t}, A_{1:t}) \\
    (2)\ &\text{Agent trust:} \ \Pr(\Tau^a_{t}\ |\ \Tau^c_{t}, Z_{1:t}, A_{1:t})
\end{aligned}
\end{equation}
remove the cross-correlations. As such, we choose to let PSMs update parameters of the conditionals and propagate those effects back to the joint posterior. 

This strategy is inspired by Gibbs sampling, a popular technique of sampling a multivariate probability density (PDF). Suppose there is a PDF, $\Pr(x,y)$ that is difficult to sample but whose marginals, $\Pr(x|y),\, \Pr(y|x)$, are easy to sample. Gibbs sampling is an iterative Markov-chain Monte Carlo method of drawing samplings representative of the joint distribution by sampling solely the conditionals.

\subsubsection{Convergence Analysis}
Gibbs sampling has formal convergence proofs for stationary distributions. However, restricting trust estimation to stationary distributions would severely handicap the model, as attacks in the real world can happen any time which will shift the true trust state. Thus, formal convergence proofs are untenable, to our knowledge. Future experimental analysis will investigate posterior estimation convergence.

\subsection{Bayesian Updating of Trust Distributions} \label{appendix:mate-bayesian}

To estimate the trust online, we perform sequential Bayesian updating. Here, we derive the general analytical form of the update. We discuss the independence assumption as presented in Section~\ref{sec:mate-update}. We then discuss the closed-form solution for our choices of prior and likelihood. 

\subsubsection{Deriving Bayesian Updating} 

We consider estimating a parameter, $\theta$ with data $X_1, X_2, ..., X_t$ arriving sequentially. In the Bayesian context, there is a prior on the parameter, $\Pr(\theta)$, and a likelihood 
\begin{equation}
\begin{aligned}
    \Pr(X_1&, X_2, ..., X_t | \theta) \coloneqq \Pr(\{X_i\}_{i=1}^t|\theta) \\
    &=\Pr(X_1|\theta) \Pr(X_2|X_1,\theta)...\Pr(X_t|X_{t-1},\theta).
\end{aligned}
\end{equation}
Finally, the posterior is
\begin{equation} \label{eq:bayesian-update-posterior}
\begin{aligned}
    \Pr(\theta | \{X_i\}_{i=1}^t) \propto \Pr(\{X_i\}_{i=1}^t|\theta) \Pr(\theta).
\end{aligned}
\end{equation}

Now, consider that we have a new data element, $X_{t+1}$. The new posterior is naturally
\begin{equation}
\begin{aligned}
    \Pr(\theta|&\{X_i\}_{i=1}^{t+1}) \propto \Pr(\{X_{i}\}_{i=1}^{t+1}|\theta) \Pr(\theta) \\
    &= \Pr(X_1|\theta) \Pr(X_2|X_1,\theta)...\Pr(X_{t+1}|X_t,\theta) \\
    &= \Pr(X_{t+1}|X_t,\theta) \Pr(\theta|\{X_i\}_{i=1}^t).
\end{aligned}
\end{equation}

Evidently, the posterior from Eq.~\ref{eq:bayesian-update-posterior} is the ``prior'' for the new data with the same likelihood function. A Bayesian update can be performed sequentially when new data arrive. 

\subsubsection{Independence Assumption} \label{appendix:mate-bayesian-independence}

Correlations may arise during the generation of agent/track PSMs since agent trust is used to compute track PSMs and track trust used to compute agent PSMs. Moreover, multiple tracks can contribute to a single agent, and vice versa. This could lead to correlation cases where e.g.,~an erroneously distrusted track makes a trusted agent appear distrusted. Unfortunately, without the independence assumption, updates to the trust posteriors using PSMs would require complex numerical methods such as Markov Chain Monte Carlo sampling because there is no appropriate parametric multi-dimensional generalization of the trust distribution - the Dirichlet is not appropriate due to its unity summation constraint. Future works will experimentally evaluate the independence assumption.

Optimal data fusion at AGG occurs when data fused from agents are uncorrelated. Agents observing the same environment from their own independent platform-level sensors will see similar state estimates but, given the environment, their tracks will importantly not contain any correlations. Tracks from each agent will exhibit possess temporal autocorrelation, but conservative fusion methods at the AGG such as from~\cite{cook1984distributed} will account for autocorrelation.

\subsubsection{Conjugate Updates in Closed Form} \label{appendix:mate-bayesian-conjugacy}
Approximating the trust as a random variable, agents and tracks are deemed either trustworthy or untrustworthy according to PSMs which follow Bernoulli distributions. A prior distribution on the Bernoulli parameter is set via a Beta distribution with hyperparameters $(\alpha,\beta)$ of the form
\begin{align*}
    \Pr(\theta | \alpha, \beta) = \frac{ \theta^{\alpha -1} (1-\theta)^{\beta-1}}{\mathcal{B}(\alpha,\beta)}
\end{align*}
where $\mathcal{B(\alpha,\beta)}$ is the Beta normalization constant. 

The posterior is then
\begin{align*}
    \Pr(\theta | \{X_i\}_{i=1}^t) &= \Pi_{i=1}^t \Pr(X_i | \theta) \Pr(\theta) \\
    &= \left(\Pi_{i=1}^t \theta^{X_i} (1-\theta)^{1-X_i} \right) \frac{ \theta^{\alpha -1} (1-\theta)^{\beta-1)}}{\mathcal{B}(\alpha,\beta)} \\
    &\propto \theta^{\sum_i^n X_i + \alpha - 1} (1-\theta)^{t - \sum_i^t + \beta - 1}.
\end{align*}
Thus the posterior is proportional to Beta($\Bar{\alpha}, \Bar{\beta}$) where $\Bar{\alpha} = \sum_i^t X_i + \alpha$ and $\Bar{\beta} = t - \sum_i^t + \beta$. 

The fact that the posterior is of the same family as the prior when the likelihood is a Bernoulli distribution means that the Beta is a conjugate prior to the Bernoulli. This often has attractive numerical and analytical properties. In this case, we can exploit the closed-form nature of the solution and perform fast updates on receipt of new PSMs.
\section{Datasets} \label{appendix:datasets}

The dataset generation pipeline from the multi-agent simulation framework in~\cite{hallyburton2023datasets} provides a guideline for generating a corpus of data from mobile and static agents. \avstack\ provides an expanded \carla\ API with python-based configuration files to set sensor and environment attributes. We generate safety-critical scenes from intersections for case study. Ground truth NPC pose is stored in global coordinates and projected into each agent's reference frame in postprocessing. The set of objects that are visible to each sensor from each agent is computed in postprocessing for fair evaluation against ground-truth. The use of multiple scenes allows us to disambiguate the impact of the algorithms from the particulars of the scene configuration. Table~\ref{tab:dataset-cases} provides insights on unique attributes of each scene. Configuration files will be released to enhance the reproducibility of our work. 

\begin{table}[t]
    \centering
    \begin{tabular}{c|c|c}
         Name & \# NPCs & Characteristics  \\
         \toprule \toprule
         Case 0 & 7 & Agent 0 occluded. \\
         Case 1 & 7 & Agent 0 occluded. Dense traffic. \\
         Case 2 & 8 & Agent 0 traverses intersection. \\
    \end{tabular}
    \caption{Baseline smart-city scenes from CARLA.}
    \label{tab:dataset-cases}
\end{table}
\section{Metrics} \label{appendix:metrics}

\subsection{Assignment metrics} \label{appendix:metrics-assignment}
Classical assignment-based metrics are commonly used to describe perception and tracking performance. Given two sets, (1) the detections/tracks (the number of positive retrievals), and (2) the truths (the number of positively labeled instances), assignment-based metrics perform a bipartite matching between the two sets to obtain three subsets: (i) the set of matches between retrievals and labels (true positives, TP), (ii) the set of lone retrievals (false positives, FP), and (iii) the set of lone labels (false negatives, FN). An assignment cost threshold is needed as a parameter to describe when a retrival is close enough to a label to be matched. We compute pairwise distance between all retrivals/labels as the cost and use the JVC algorithm~\cite{1986blackmanRadar} to obtain assignments with the cost threshold set at $2~m$. Statistics over the assignment outcome including precision, recall, and the F1-score as the harmonic mean between precision and recall are used to summarize the performance of an algorithm. These metrics are defined as

\begin{equation} \label{eq:assignment-metrics}
    \begin{aligned}
        \textbf{Precision:} \quad & \frac{\#\text{TP}}{\#\text{TP} + \#\text{FP}}\\
        \textbf{Recall:} \quad & \frac{\#\text{TP}}{\#\text{TP} + \#\text{FN}} \\
        \textbf{F1 Score:} \quad & \frac{2 \ \text{Precision} \cdot \text{Recall}}{\text{Precision} + \text{Recall}}.
    \end{aligned}
\end{equation}

An adversary injecting false positives will deteriorate the precision score while an adversary inclined to false negative attacks will compromise the recall. A translation-based attacker will affect both measures of performance.

\subsection{OSPA metrics} \label{appendix:metrics-ospa}
The Optimal SubPattern Assignment metric is a mixing between assignment cost (state estimation error, i.e., the distance between a track assigned to a truth), a penalty for unassigned tracks/truths, and a penalty for different numbers of tracks/truths. Given a bipartite assignment between the set of tracks and the set of truths, $\mathcal{A} = \{a_{i,\text{track}},\, a_{i,\text{truth}}\}_{i=1}^{n_a}$, with $n_a = |\mathcal{A}|$ the number of assignments, the \texttt{cost} of an assignment as $\texttt{dist}(a_{i,\text{track}},\,a_{i,\text{truth}})$, and the unassigned tracks/truths as $\Bar{\mathcal{A}}$, OSPA is
\begin{equation} \label{eq:ospa-metric}
    \begin{aligned}
        \textbf{OSPA} \coloneqq \frac{1}{n} \left( \sum_{i}^{n_a} \mathcal{A}_i.\texttt{cost}() + c \, |\mathcal{\Bar{\mathcal{A}}}| \right)^p + c^p (n - m)
    \end{aligned}
\end{equation}
with $c$ and $p$ user-defined weighting parameters, $n$ and $m$ the smaller and larger of the number of tracks and truths, respectively, and $|\mathcal{\Bar{\mathcal{A}}}|$ the smaller of the number of unassigned tracks and unassigned truths. Lower OSPA implies lower cost/penalty and is desirable. For a full derivation and proofs on OSPA as a proper metric, see~\cite{schuhmacher2008ospa}.

\subsection{Novel trust metrics} \label{appendix:metrics-trust}
To measure trust estimation performance, we build novel trust-oriented metrics for both track and agent trust. Both use the distance between a reference probability density (PDF), $f_r(\tau)$, (i.e., the estimated trust) and a target PDF, $f_t(\tau)$, (i.e., the true trust). The distance is the area between the reference and target cumulative distribution functions (CDFs), $F_r(\tau),\, F_t(\tau)$, i.e.,
\begin{equation}
    \begin{aligned}
        D_{\tau} &= \int_0^1 |F_t(\tau) - F_r(\tau)| d\tau.
    \end{aligned}
\end{equation}
For a true trust that is binary (completely trusted or distrusted), the target CDF is either $F_t(\tau) = 1.0$ for false/distrusted or $F_t(\tau) = 0.0 \ \texttt{if} \ \tau < 1 \ \texttt{else} \ 1.0$. Thus, for a false/distrusted target, $F_t(\tau)$ majorizes $F_r(\tau)$ while for a true/trusted target, $F_t(\tau)$ minorizes $F_r(\tau)$. This allows us to remove the absolute value in different target cases to simplify the integration as
\begin{equation}
    \begin{aligned}
        D_{\tau} &= \begin{cases}
            1 - \int_0^1 F_r(\tau) d\tau & \text{target is false/distrusted} \\
            \int_0^1 F_r(\tau) d\tau & \text{target is true/trusted}
        \end{cases}
    \end{aligned}
\end{equation}
Moreover, with integration by parts on any CDF $F_X(x)$ whose PDF $f(x)$ has support on $[a,b]$ (i.e.,~$F_X(a)=0,\, F_X(b)=1$),
\begin{equation}
    \begin{aligned}
        \int_a^b F_X(x) dx &= \left[x F_X(x)\right]_a^b - \int_a^b F'_X(x) dx \\
        &= (bF_X(b) - aF_X(a)) - \int_a^b x f(x) dx \\
        &= b - \expectation[f(x)]
    \end{aligned}
\end{equation}
As a result, the distance simplifies to
\begin{equation} \label{eq:trust-distance}
    \begin{aligned}
        D_{\tau} &= \begin{cases}
            \expectation[f_r(\tau)] & \text{target is false/distrusted} \\
            1 - \expectation[f_r(\tau)] & \text{target is true/trusted.}
        \end{cases}
    \end{aligned}
\end{equation}

The remaining consideration for track and agent trust distances ($D_{\tau}^{\text{track}},\, D_{\tau}^{\text{agent}})$ is to determine the target distribution class (false/distrusted vs. true/trusted). 

\paragraph{Track trust metric, $D_{\tau}^{\text{track}}$.} For each track state with its associated reference trust distribution, it is simple to determine the target distribution. Given the set of true objects, a track has target of \texttt{true} if the bipartite matching (assignment) algorithm between all tracks and all truths with suitable threshold distance yields a positive assignment for the track in question and \texttt{false} otherwise.

\paragraph{Agent trust metric, $D_{\tau}^{\text{agent}}$.} For each agent with its associated reference agent trust distribution, the determination of the target distribution is more nuanced. Agents will be maintaining a collection of locally tracked objects each representing either a true object or a false object. Natural false tracks may occur from time to time, however, the overwhelming majority of the tracks should represent true objects. To determine the target trust for the agent, we perform an assignment of all agent's local tracks to the locally-viewable true objects and obtain the tracking F1-score (see Appendix~\ref{appendix:metrics-assignment} above). We set a trustworthy F1-score threshold, $\mathcal{T}_{\texttt{F1}} $ meaning,
\begin{align}
    \text{Agent Reference} = \begin{cases}
         \texttt{trusted} & \texttt{F1} > \mathcal{T}_{\texttt{F1}} \\
        \texttt{distrusted} & \text{otherwise}
    \end{cases}.
\end{align}
An alternative approach to agent trust target determination is to maintain an oracle set of the agents that are attacked and to assign a target of \texttt{distrusted} to agents in that set and \texttt{trusted} to all other agents.
\section{Additional Results Insights} \label{appendix:results}

\subsection{Monte Carlo model tuning} \label{appendix:results-mc-tuning}

We randomize trust model parameters according to the random functions in Table~\ref{tab:trust-mc} allowing us to tune the trust model on the available Monte Carlo adversary datasets. We run several thousand versions of security-aware sensor fusion on top of the hundreds of adversarial datasets and capture assignment, OSPA, and trust metrics at each frame for each algorithm instantiation. 


\begin{table}[t]
    \centering
    \begin{tabular}{c|c|c}
        \textbf{Parameter} & \textbf{Function} & \textbf{Inputs} \\
        \toprule \toprule
        $\propagatorprioralpha$ (Eq.~\ref{eq:trust-propagator-prior}) & \texttt{RANDN} & $\expectation[\alpha_0]=\mu_{\alpha_0}$, $\variance[\alpha_0]=\sigma_{\alpha_0}^2$ \\
        $\propagatorpriorbeta$ (Eq.~\ref{eq:trust-propagator-prior}) & \texttt{RANDN} & $\expectation[\beta_0]=\mu_{\beta_0}$, $\variance[\beta_0]=\sigma_{\beta_0}^2$ \\
        $\agentnegativitybias$ (Eq.~\ref{eq:beta-bernoulli-update-weighted}) & \texttt{RAND} & $[1,\ 20]$\\
        $\tracknegativitybias$ (Eq.~\ref{eq:beta-bernoulli-update-weighted}) & \texttt{RAND} & $[1,\ 10]$\\
        $\agentnegativitythreshold$ (Eq.~\ref{eq:beta-bernoulli-update-weighted})& \texttt{RAND} & $[0,\ 0.6]$ \\
        $\tracknegativitythreshold$ (Eq.~\ref{eq:beta-bernoulli-update-weighted}) & \texttt{RAND} & $[0,\ 0.6]$\\
        $\thresholdtrackignore$ (Sec.~\ref{sec:trusted-fusion}) & \texttt{RAND} & $[0.2,\ 0.8]$ \\
        $\weightedkalmanupdate$ (Sec.~\ref{sec:trusted-fusion}) & \texttt{LOGRAND} & $[0.1,\ 10]$
        \end{tabular}
    \caption{A subset of the randomizable parameters for Monte Carlo trust algorithm tuning. Parameters are motivated in their corresponding equations. The randomized tuning couple with metrics from Sec.~\ref{sec:exp-ue-metrics} gives objective algorithm tuning on a training dataset.}
    \label{tab:trust-mc}
\end{table}

\subsection{Case study visualizations} \label{appendix:results-case-studies}

\begin{figure*}[t]
    \centering
    \begin{multicols}{2}
    \foreach \x in {0,1,2,3}
    {
        \begin{subfigure}[b]{0.8\linewidth}
            \centering
            \includegraphics[width=0.95\linewidth,trim={2cm 4.5cm 2cm 4.5cm},clip,fbox]{diagrams/ros_scene_visualizations/case0/case0_agent\x_benign.png}
            \caption{Benign, Agent \x}
        \end{subfigure}
        \vspace{12pt}
    }
    \begin{subfigure}[b]{0.8\linewidth}
        \centering
        \includegraphics[width=0.95\linewidth,trim={2cm 4.5cm 2cm 4.5cm},clip,fbox]{diagrams/ros_scene_visualizations/case0/case0_cc_benign.png}
        \caption{Benign, AGG}
    \end{subfigure}
    \vspace{12pt}
    \begin{subfigure}[b]{1.3\linewidth}
        \centering
        \includegraphics[width=0.6\linewidth,trim={2cm 4.5cm 2cm 4.5cm},clip,fbox]{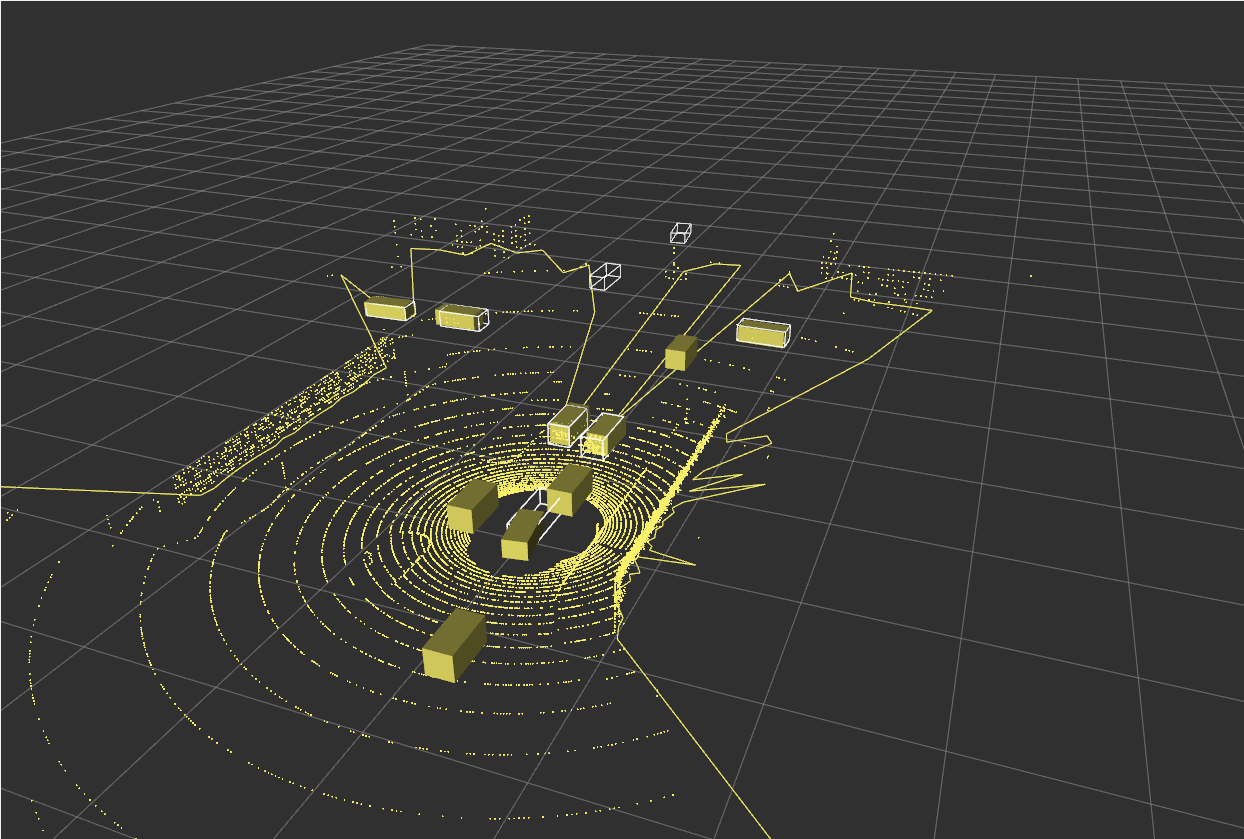}
        \caption{FP attack on Agent 0}
    \end{subfigure}
    \begin{subfigure}[b]{1.3\linewidth}
        \centering
        \vspace{-12pt}
        \includegraphics[width=0.6\linewidth,trim={2cm 4.5cm 2cm 4.5cm},clip,fbox]{diagrams/ros_scene_visualizations/case0/case0_cc_attacked.png}
        \caption{Non-trust-aware AGG compromised!}
    \end{subfigure}
    \begin{subfigure}[b]{1.2\linewidth}
        \hspace{-10mm}
        \includegraphics[width=0.98\linewidth,trim={0cm 0cm 0cm 0cm},clip,fbox]{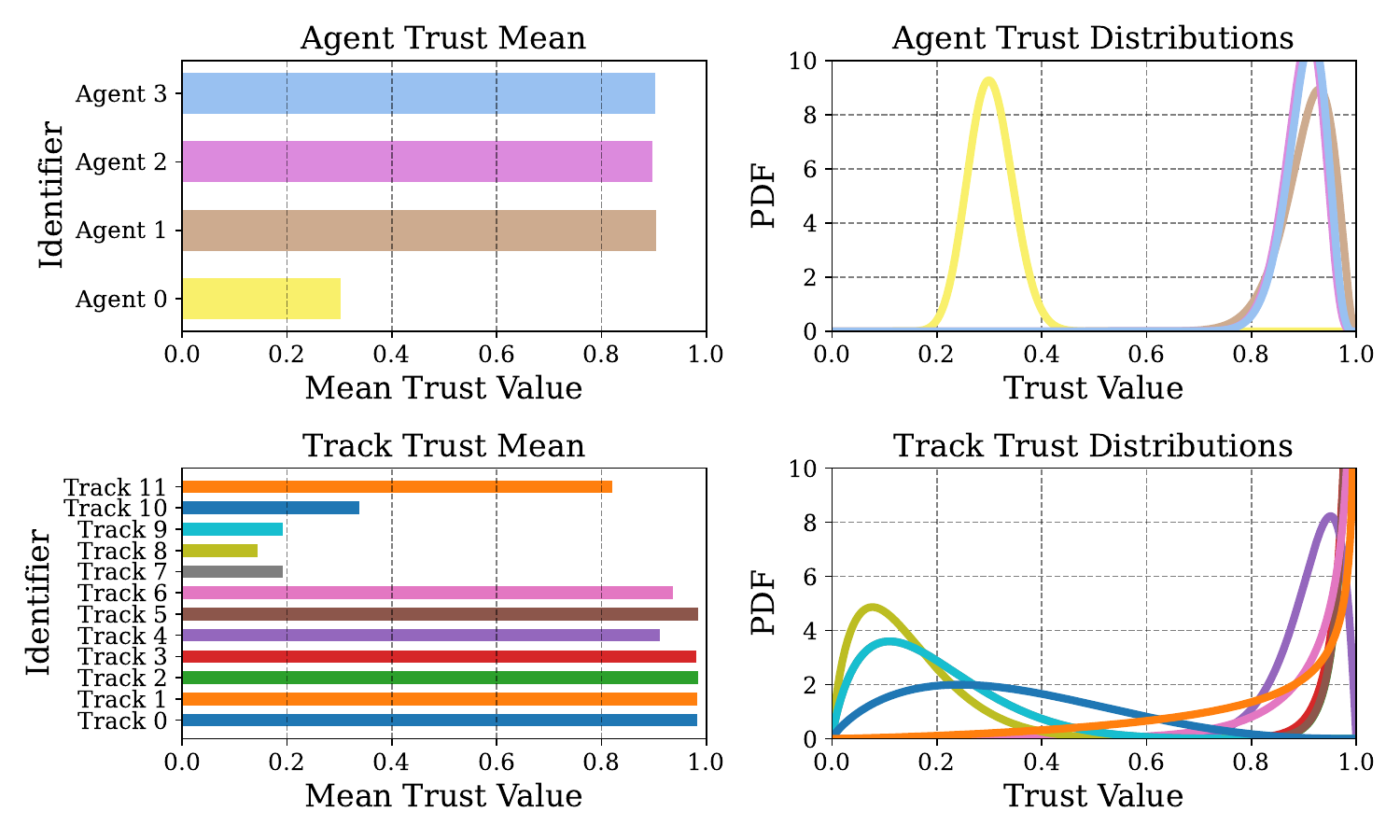}
        \caption{Trust outcomes for attacked scenario}
    \end{subfigure}
    \hspace{-10mm}
    \begin{subfigure}[b]{1.2\linewidth}
        \hspace{-10mm}
        \includegraphics[width=0.98\linewidth,trim={0cm 0cm 0cm 0cm},clip,fbox]{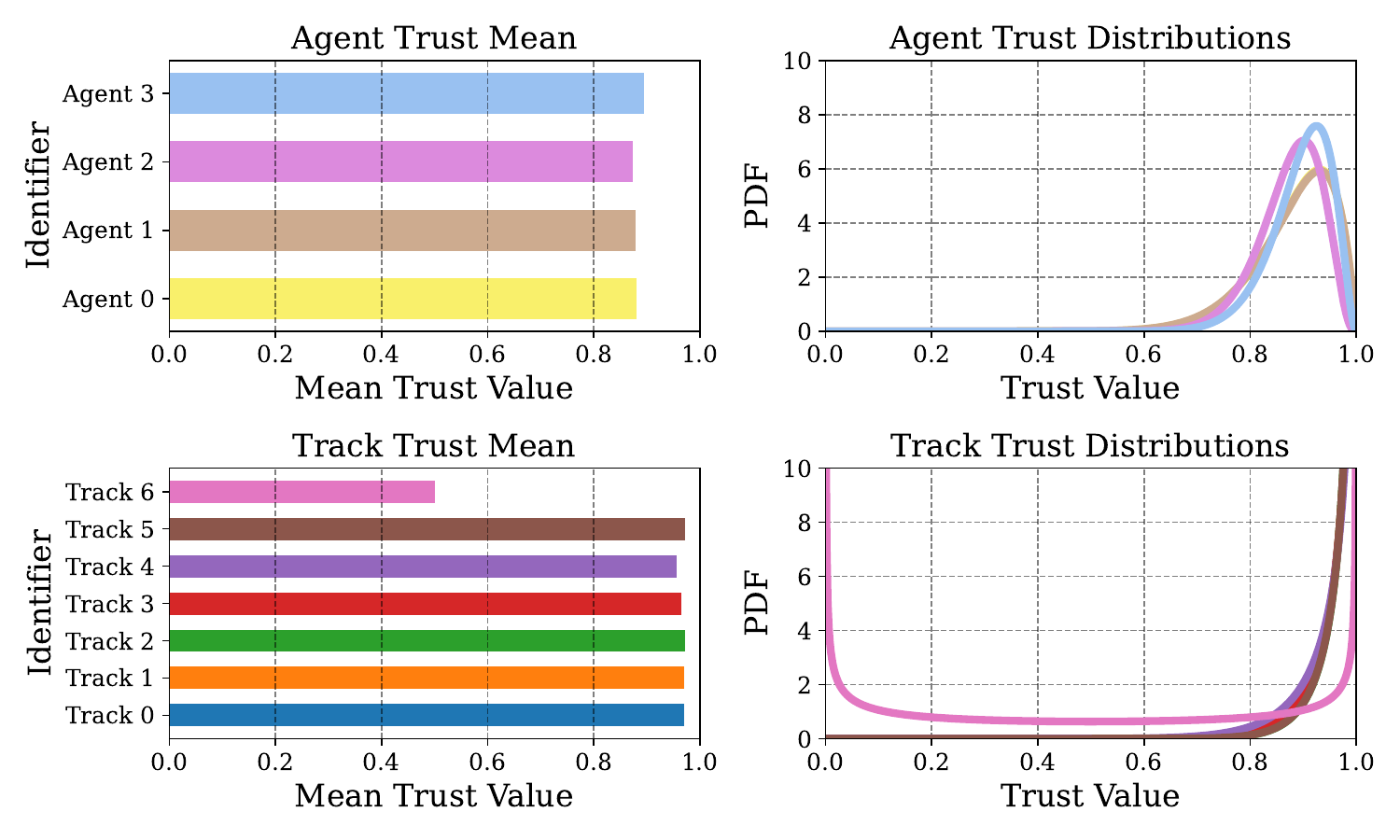}
        \caption{Trust outcomes for benign baseline scenario}
    \end{subfigure}
    \end{multicols}
    \vspace{-12pt}
    \caption{(a-d) In case 0 benign baseline, agent tracking fused at (e) AGG yields (i) high trust on agents and tracks. (f) Attacked Agent 0 leads to (g) false tracks detected with (h) trust estimation in security-aware sensor fusion. (i) Security-aware fusion does not compromise the unattacked baseline.}
    \label{fig:results-viz-case-0}
\end{figure*}

\begin{figure*}[t]
    \centering
    \begin{multicols}{2}
    \foreach \x in {0,1,2,3}
    {
        \begin{subfigure}[b]{0.8\linewidth}
            \centering
            \includegraphics[width=0.95\linewidth,trim={2cm 4.5cm 2cm 4.5cm},clip,fbox]{diagrams/ros_scene_visualizations/case1/case1_agent\x_benign.png}
            \caption{Benign, Agent \x}
        \end{subfigure}
        \vspace{12pt}
    }
    \begin{subfigure}[b]{0.8\linewidth}
        \centering
        \includegraphics[width=0.95\linewidth,trim={2cm 4.5cm 2cm 4.5cm},clip,fbox]{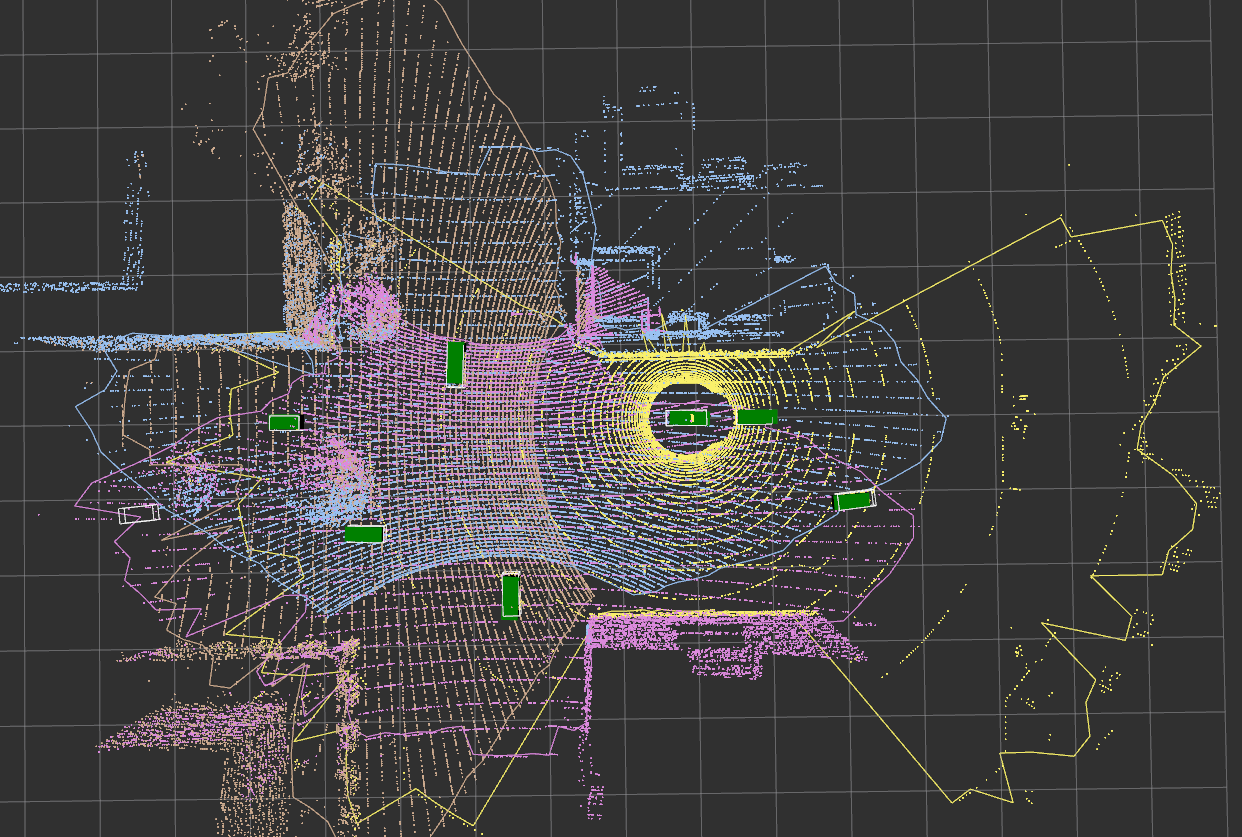}
        \caption{AGG}
    \end{subfigure}
    \vspace{12pt}
    \begin{subfigure}[b]{1.3\linewidth}
        \centering
        \includegraphics[width=0.6\linewidth,trim={2cm 4.5cm 2cm 4.5cm},clip,fbox]{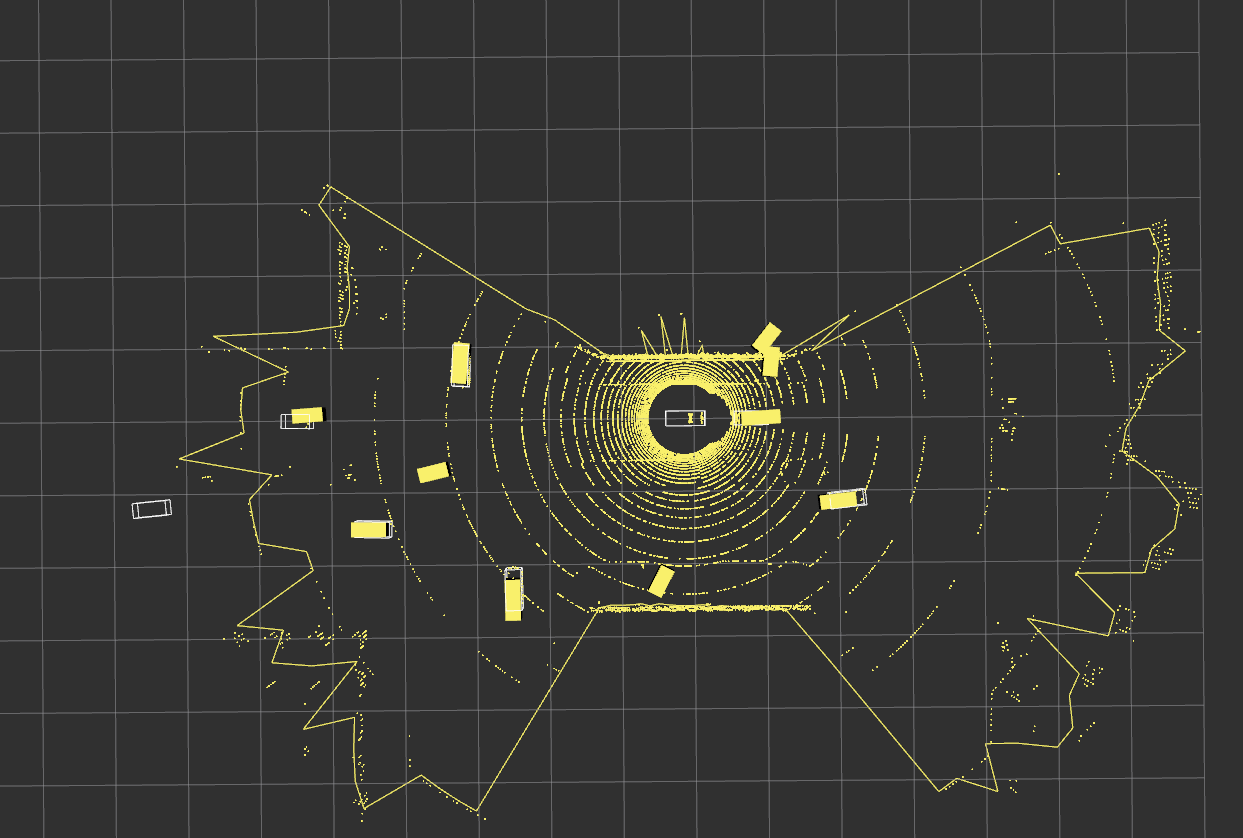}
        \caption{FP attack on Agent 0}
    \end{subfigure}
    \begin{subfigure}[b]{1.3\linewidth}
        \centering
        \vspace{-12pt}
        \includegraphics[width=0.6\linewidth,trim={2cm 4.5cm 2cm 4.5cm},clip,fbox]{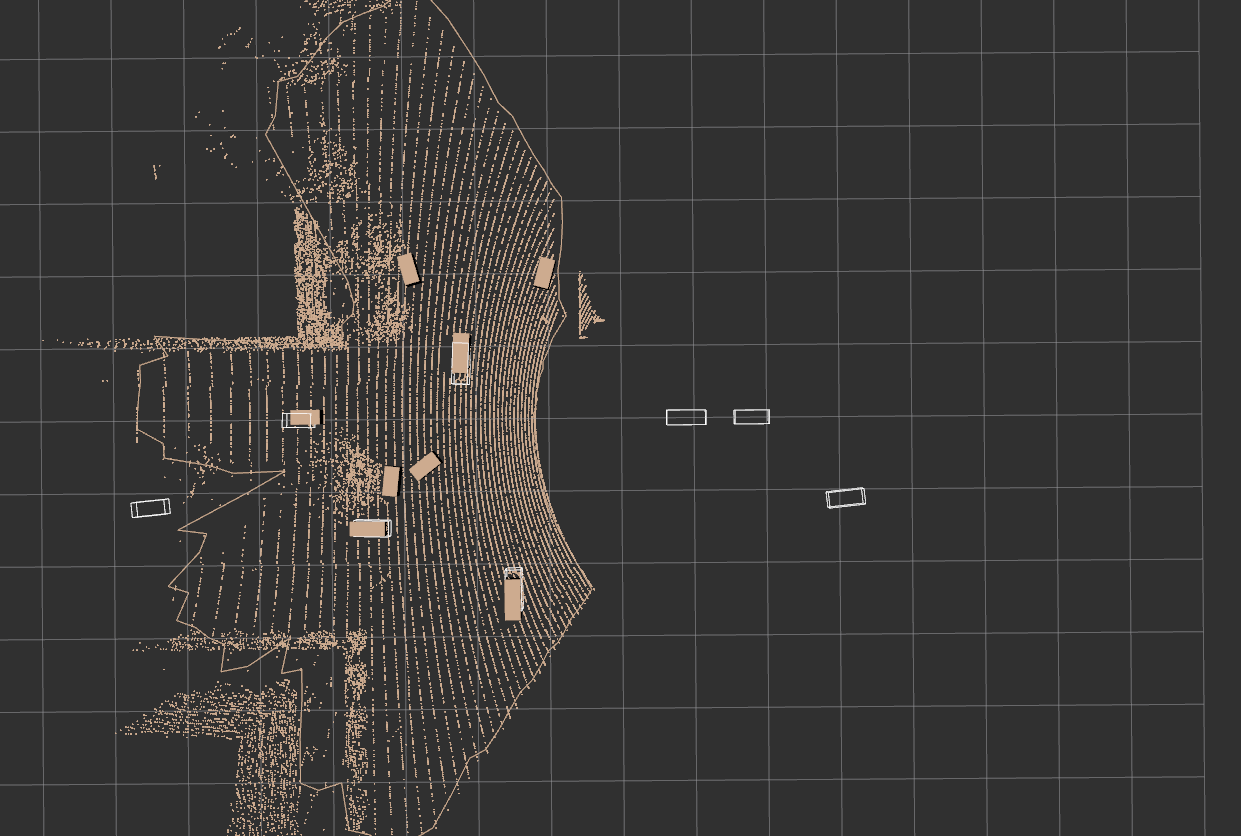}
        \caption{FP attack on Agent 1}
    \end{subfigure}
    \begin{subfigure}[b]{1.2\linewidth}
        \hspace{-10mm}
        \includegraphics[width=0.98\linewidth,trim={0cm 0cm 0cm 0cm},clip,fbox]{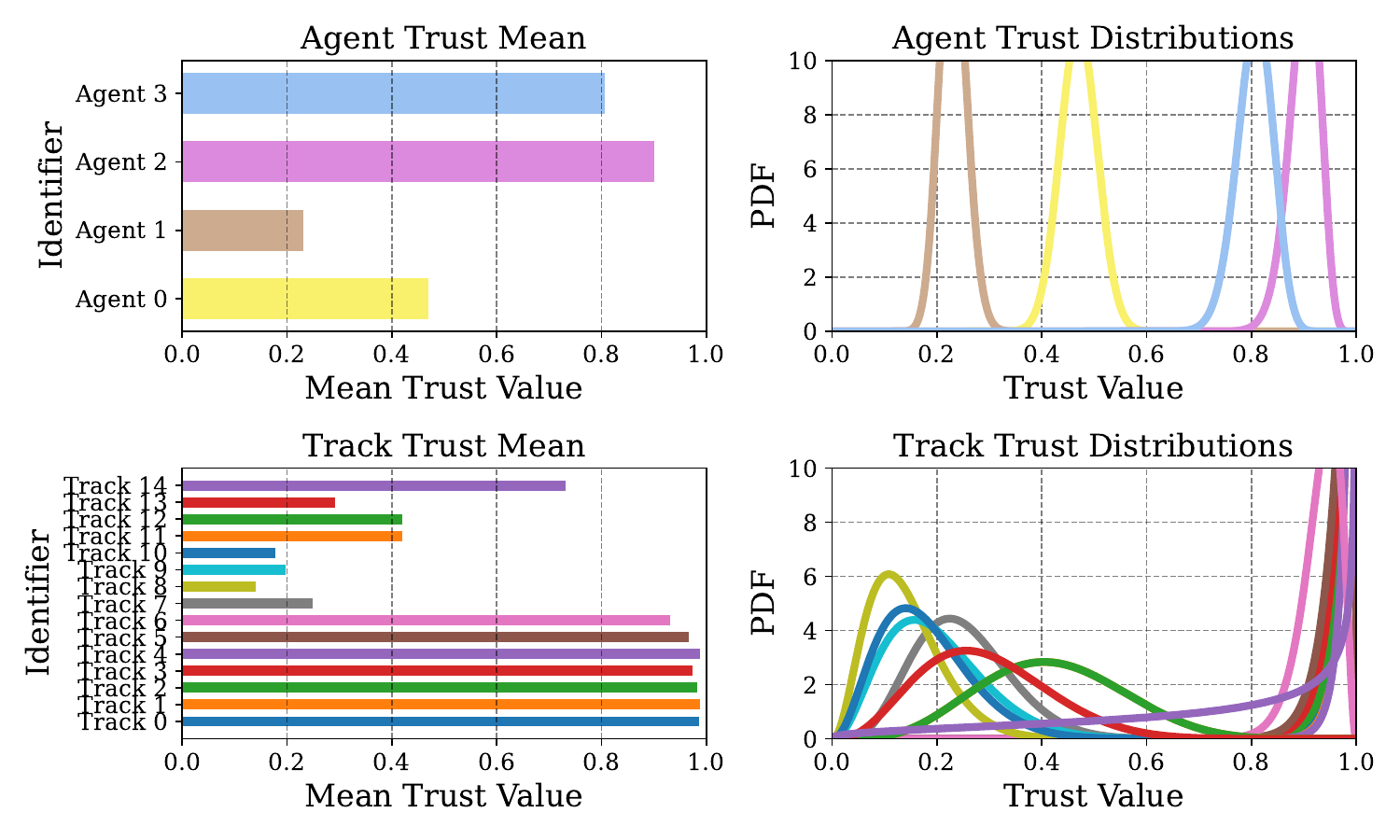}
        \caption{Trust outcomes for attacked scenario}
    \end{subfigure}
    \hspace{-10mm}
    \begin{subfigure}[b]{1.2\linewidth}
        \hspace{-10mm}
        \includegraphics[width=0.98\linewidth,trim={0cm 0cm 0cm 0cm},clip,fbox]{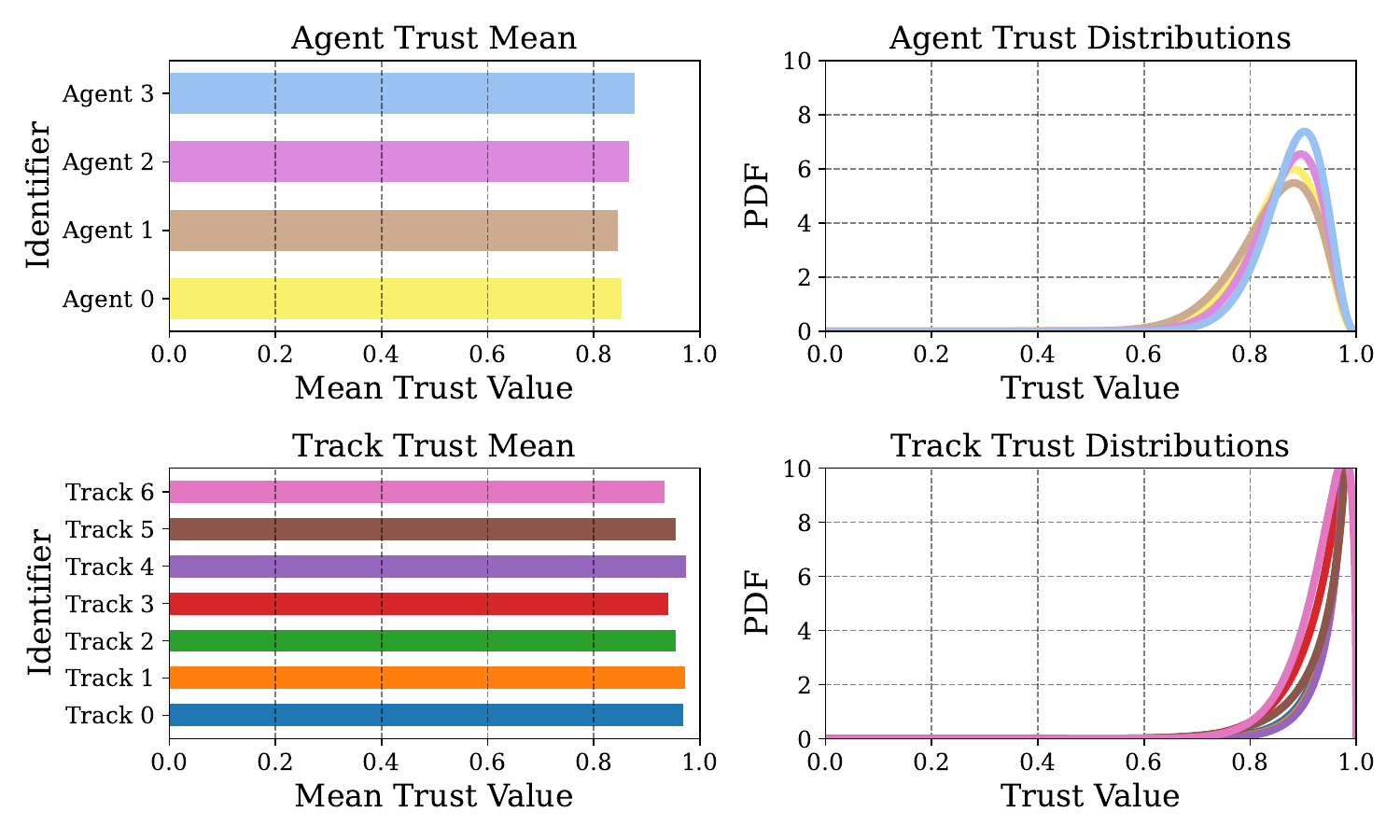}
        \caption{Trust outcomes for benign baseline scenario}
    \end{subfigure}
    \end{multicols}
    \vspace{-12pt}
    \caption{(a-d) In case 1 benign baseline, agent tracking fused at (e) aggregator yields (i) high trust on agents and tracks. (f-g) Agents 0, 1 compromised by FP attacker (h) that is detected by security-aware sensor fusion. (i) Security-aware fusion does not compromise the unattacked baseline.}
    \label{fig:results-viz-case-1}
\end{figure*}

\begin{figure*}[t]
    \centering
    \begin{multicols}{2}
    \foreach \x in {0,1,2,3}
    {
        \begin{subfigure}[b]{0.8\linewidth}
            \centering
            \includegraphics[width=0.95\linewidth,trim={2cm 4.5cm 2cm 4.5cm},clip,fbox]{diagrams/ros_scene_visualizations/case2/case2_agent\x_benign.png}
            \caption{Benign, Agent \x}
        \end{subfigure}
        \vspace{12pt}
    }
    \begin{subfigure}[b]{0.8\linewidth}
        \centering
        \includegraphics[width=0.95\linewidth,trim={2cm 4.5cm 2cm 4.5cm},clip,fbox]{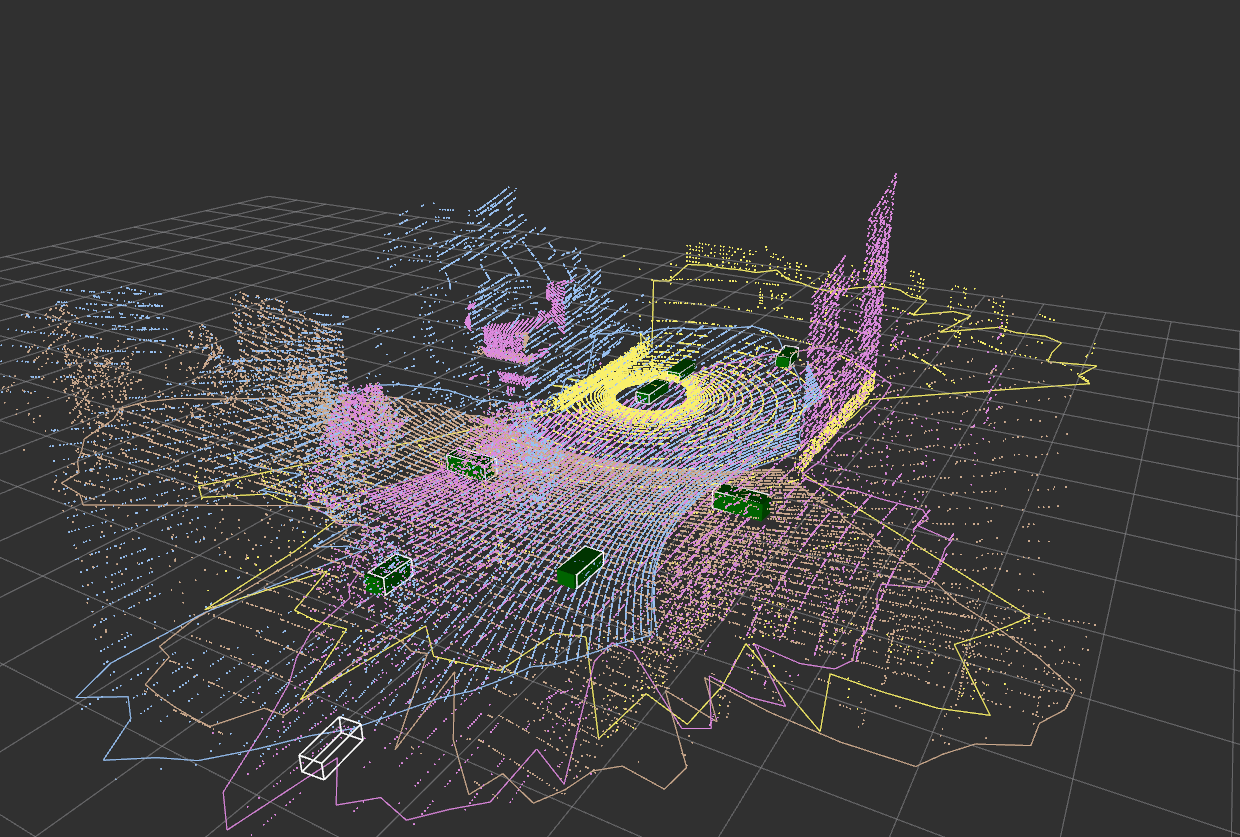}
        \caption{Benign, AGG}
    \end{subfigure}
    \vspace{12pt}
    \begin{subfigure}[b]{1.3\linewidth}
        \centering
        \includegraphics[width=0.6\linewidth,trim={2cm 4.5cm 2cm 4.5cm},clip,fbox]{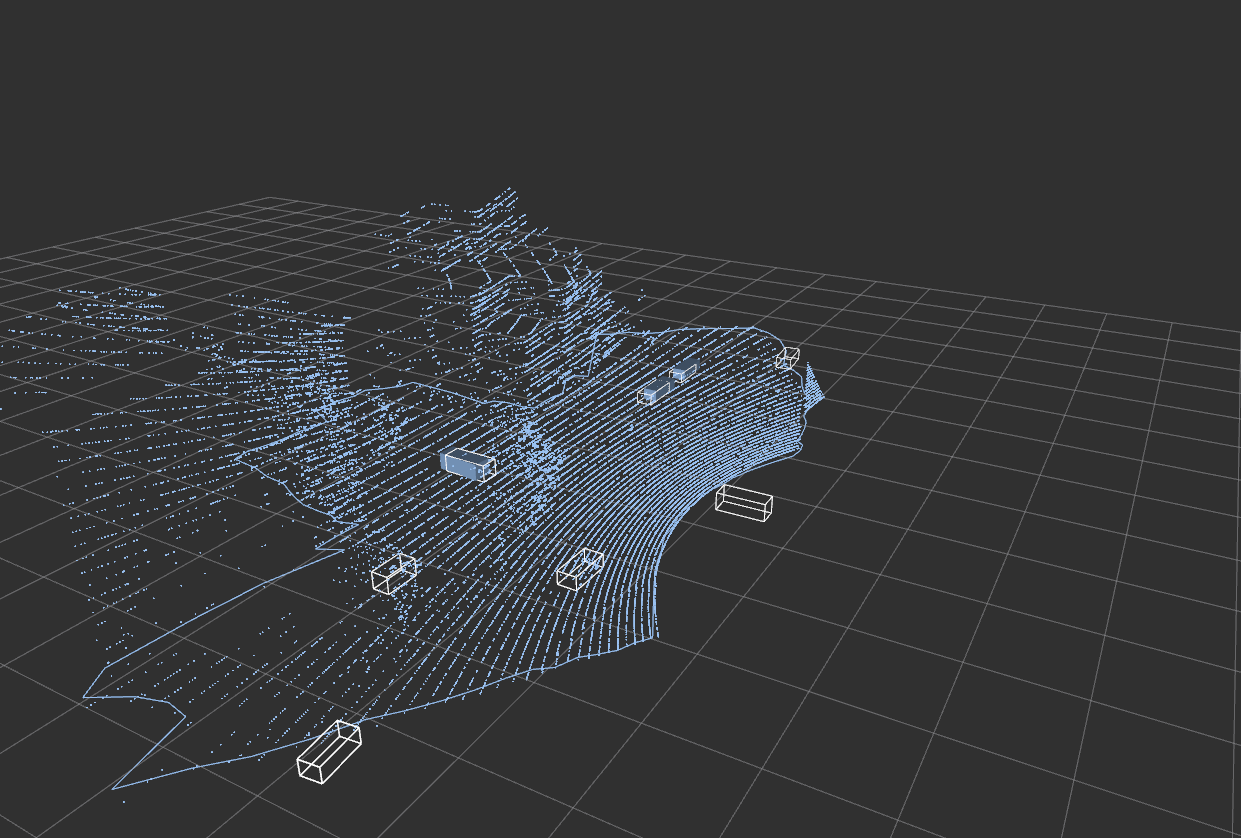}
        \caption{FN attack on agent3}
    \end{subfigure}
    \begin{subfigure}[b]{1.3\linewidth}
        \centering
        \vspace{-12pt}
        \includegraphics[width=0.6\linewidth,trim={2cm 4.5cm 2cm 4.5cm},clip,fbox]{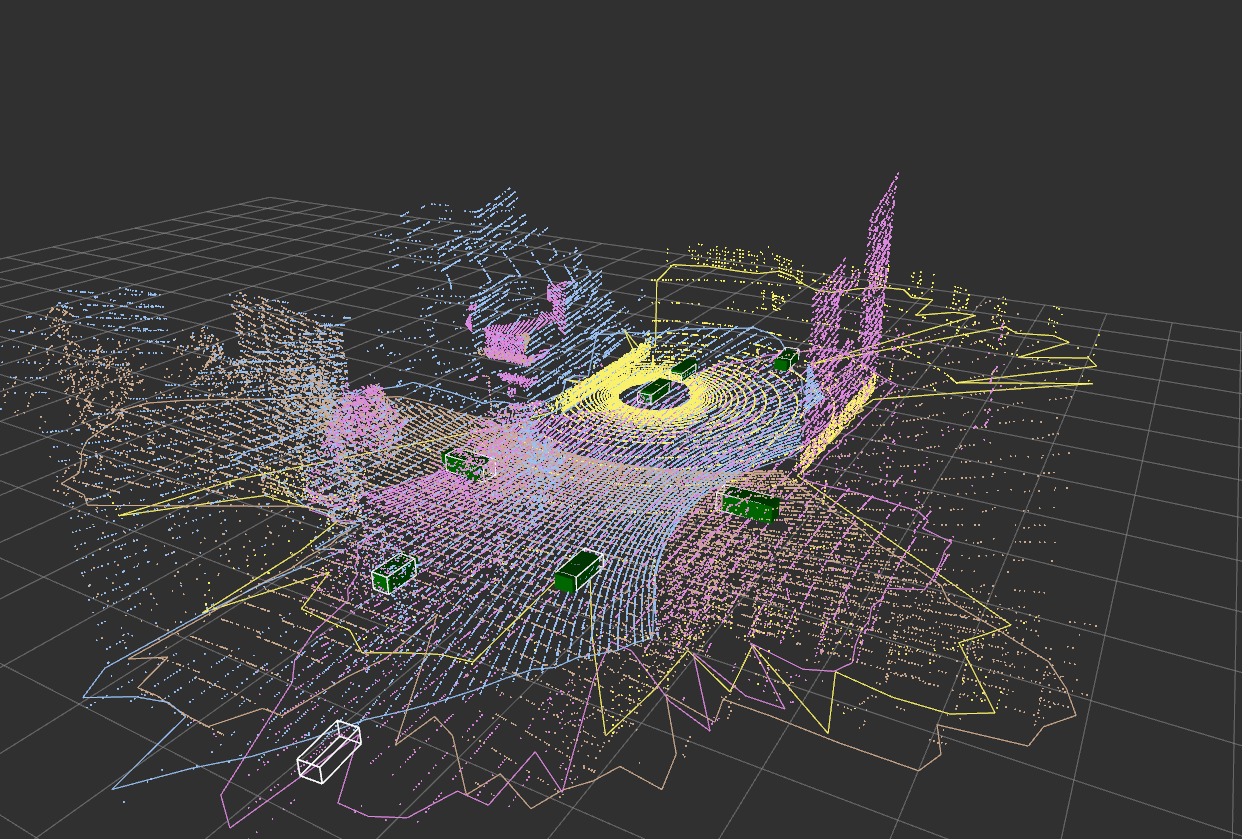}
        \caption{Other agents compensate for FN attack}
    \end{subfigure}
    \begin{subfigure}[b]{1.2\linewidth}
        \hspace{-10mm}
        \includegraphics[width=0.98\linewidth,trim={0cm 0cm 0cm 0cm},clip,fbox]{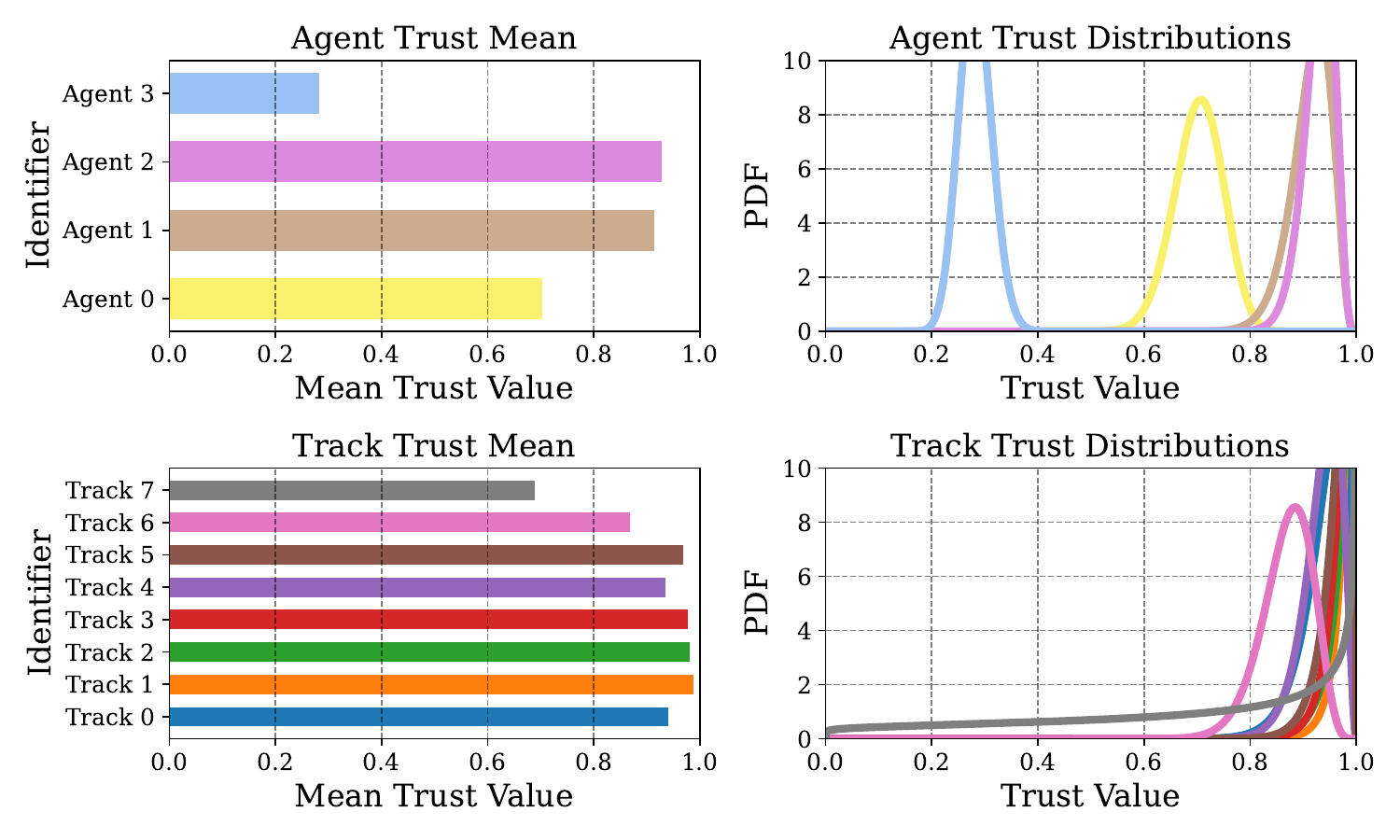}
        \caption{Trust outcomes for attacked scenario}
    \end{subfigure}
    \hspace{-10mm}
    \begin{subfigure}[b]{1.2\linewidth}
        \hspace{-10mm}
        \includegraphics[width=0.98\linewidth,trim={0cm 0cm 0cm 0cm},clip,fbox]{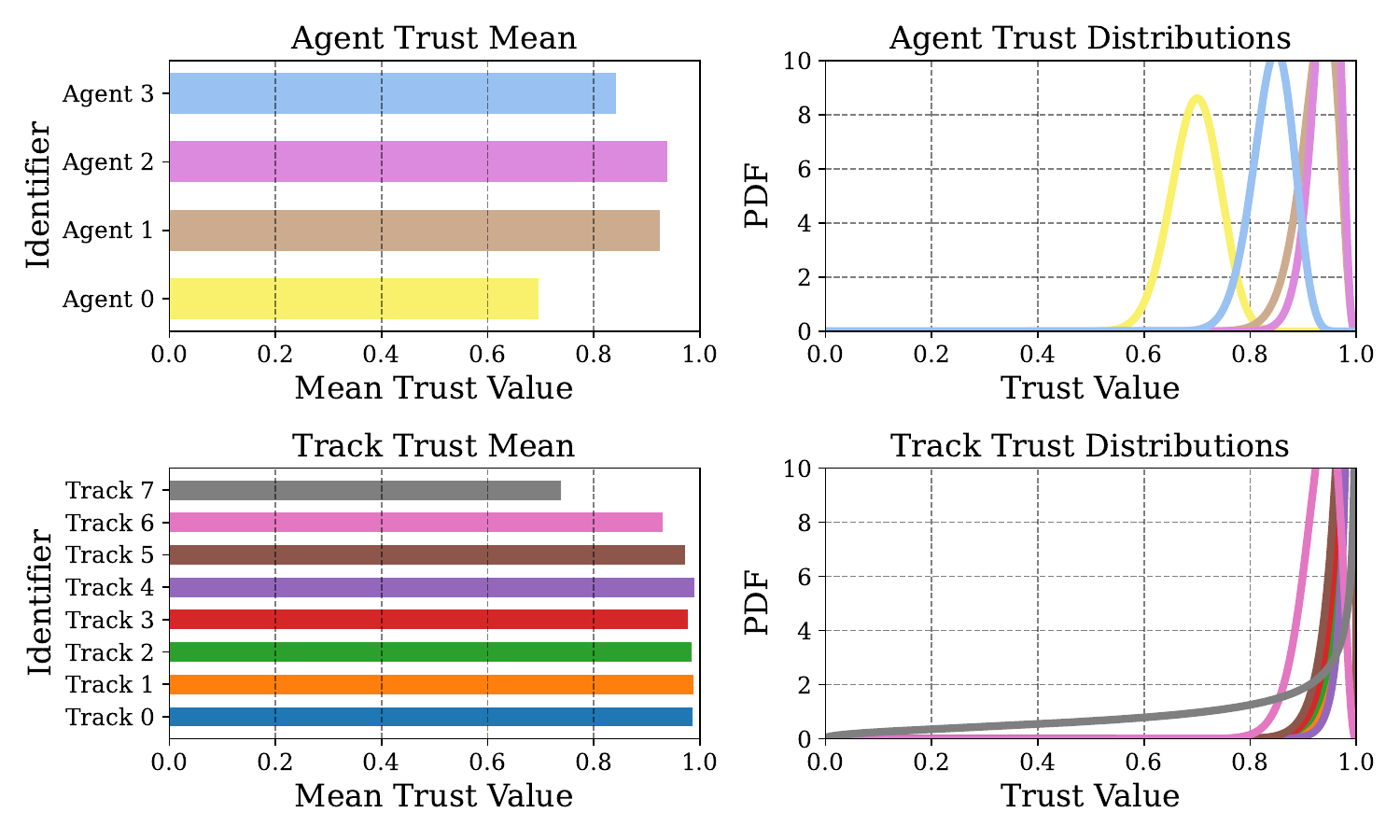}
        \caption{Trust outcomes for benign baseline scenario}
    \end{subfigure}
    \end{multicols}
    \vspace{-12pt}
    \caption{(a-d) In case 2 benign baseline, agent tracking fused at (e) aggregator yields (i) high trust on agents and tracks. (f) FN attack on Agent 3 (h) is detected quickly by trust estimation without compromising existing valid track trust. (i) Security-aware fusion does not compromise the unattacked baseline.}
    \label{fig:results-viz-case-2}
\end{figure*}

Section~\ref{sec:exp-ue-dataset-case-results} presented the results for the three in-depth case studies. Shown previous were snapshots of the compromised agents, the estimated trust distributions, and performance metrics using the assignment, OSPA, and trust-based metrics. Additional visualizations for each of the agents for each of the cases as well as the full agent and track trust distributions are presented here in Figures~\ref{fig:results-viz-case-0},~\ref{fig:results-viz-case-1}, and~\ref{fig:results-viz-case-2}.

\end{document}